\documentclass[a4paper,fleqn]{cas-sc}

\usepackage{caption}
\usepackage[utf8]{inputenc}
\usepackage{float}  
\usepackage{lastpage}
\usepackage{fancyhdr}
\usepackage{hyperref}

\usepackage[authoryear]{natbib}

\def\tsc#1{\csdef{#1}{\textsc{\lowercase{#1}}\xspace}}
\tsc{WGM}
\tsc{QE}
\tsc{EP}
\tsc{PMS}
\tsc{BEC}
\tsc{DE}

\begin{document}
\let\WriteBookmarks\relax
\def\floatpagepagefraction{1}
\def\textpagefraction{.001}
\shorttitle{ }

\title{Quantifying Interface–Procedure Coupling Risks in Digital Nuclear Control Rooms: An Event-Based Human Reliability Assessment}     

\tnotetext[1]{The research was supported by a grant from the National Natural Science Foundation of China (Grant No. T2192933),  the LingChuang Research Project of China National Nuclear Corporation, the Foundation of National Key Laboratory of Human Factors Engineering (Grant No. HFNKL2024W07), and Tsinghua University Initiative Scientific Research Program.}

\author[1,2]{Xingyu Xiao}[style=chinese]
\credit{Conceptualization, Methodology, Software, Formal analysis, Data Curation, Visualization, Validation, Writing- Original draft preparation.}

\author[3]{Mingwei Xia}[style=chinese]
\credit{Resources.}

\author[3]{Hongbo Li}
\credit{Resources.}

\author[1]{Jingang Liang}[orcid=0000-0003-2632-8613]
\cortext[cor1]{Corresponding author}
\cormark[1]
\credit{Formal analysis.}

\author[1,2]{Jiejuan Tong}[style=chinese]
\credit{ Formal analysis.}

\author[1]{Haitao Wang}[style=chinese]
\credit{ Formal analysis.}

\affiliation[1]{organization={Institute of Nuclear and New Energy Technology, Tsinghua University},
            city={Beijing},
            postcode={100084}, 
            country={China}}
\affiliation[2]{organization={National Key Laboratory of Human Factors Engineering},
            city={Beijing},
            postcode={100094}, 
            country={China}}
\affiliation[3]{organization={China Nuclear Power Operation Technology Corporation,ltd.},
city={Wuhan},
postcode={430223}, 
country={China}}

\begin{abstract}
Digitalization has reshaped human–system interaction in nuclear main control rooms, yet the quantitative mechanisms through which interfaces amplify procedural risks remain largely undocumented. This study provides a systematic quantitative assessment of interface–procedure coupling in a modern nuclear power plant, revealing how digital interface semantics fundamentally alter operators’ cognitive pathways. Using real operational events collected between 2021 and 2025, we develop a reusable three-dimensional labeling scheme (procedure, interface, coupling) together with a four-factor UI mechanism model that structurally decomposes layout, semantic, mismatch, and labeling issues. Our analyses uncover that UI problems act as a risk amplifier: 42.6\% of all events involved interface deficiencies, and interface involvement more than doubles the likelihood of procedural deviation (OR = 2.35). Machine-learning interpretation further identifies the composite PSF of interface–procedure coupling, specifically driven by semantic mismatches and layout traps, as the dominant driver of coupled failures, marking a shift from traditional rule-following errors toward interface-induced cognitive integration failures in digital MCRs. Experimental validation with simulator-based parameter verification tasks confirms that semantic confusion accounts for 27.3\% of interface-induced errors, with an overall error rate matching historical event patterns. Beyond empirical insights, this work delivers a data-driven HRA workflow tailored for supporting the early identification of vulnerabilities in digitalized control rooms and proposes a systematic interface–procedure semantic alignment framework for informing risk-aware design and verification. Together, these contributions establish a foundational quantitative basis for characterizing, assessing, and mitigating interface–procedure coupling traps in next-generation nuclear control rooms. 
\end{abstract}

\begin{keywords}
Human reliability analysis\sep Digital control room\sep Interface–procedure coupling\sep Semantic misalignment\sep Machine Learning\sep Human factors \sep Composite PSF

\end{keywords}

\maketitle

\section{Introduction}\label{Introduction}
The digitalization of nuclear power plant (NPP) main control rooms (MCRs) has fundamentally transformed the nature of human–system interaction (HSI) \cite{liu2017identifying}. Modern digital MCRs integrate advanced monitoring displays, soft controls, and computer-based procedures (CBPs) \cite{oxstrand2014computer}. While these technologies provide operators with enhanced decision-support functionality, they also introduce cognitive complexity and semantic ambiguity \cite{qi2022open}. As interfaces become increasingly multi-layered, the alignment between interface semantics and procedural logic is no longer guaranteed. Under these conditions, small inconsistencies can propagate into operationally significant errors \cite{xiao2024emergency}.

Historically, studies of human error in nuclear operations emphasized procedure-following behavior and execution slips \cite{xiao2025dynamic}. First-generation and second-generation human reliability analysis (HRA) frameworks typically treated procedures and interfaces as static contextual factors \cite{xiao2025comprehensive}. In contrast, third-generation frameworks acknowledge that operators must integrate heterogeneous cues across multiple digital displays \cite{xiao2025autograph}. However, empirical evidence quantifying how digital interface design interacts with procedural workflows remains limited. Most existing studies rely on laboratory experiments or qualitative analyses rather than real-world operational data \cite{tan2026dynamic}.

Operational experience indicates a growing class of failures where interface deficiencies amplify procedural deviations \cite{chang1999development}. These "coupling traps" arise when procedural steps assume interface states that are inconsistent with operator perception \cite{yan2017effect}. Mismatches in layout or status indication can misguide operator cognition, leading to incorrect action sequences even when procedures are technically correct \cite{xing2024functional}. Such emergent risks are often overlooked by traditional HRA methods, which treat procedure and interface errors as separate categories rather than mutually reinforcing mechanisms \cite{park2025dynamic}.

To address this gap, the present study provides a data-driven quantification of \textbf{interface-procedure coupling} using multi-year operational event records. We define this coupling as the dynamic interdependence between procedural logic and interface semantics. Specifically, we analyze 59 A-class and B-class events (2021–2025) using a structured labeling scheme to classify procedural deviations and interface deficiencies. Interface problems are further decomposed into semantic, spatial, and labeling factors. Statistical analyses, combined with a random forest model, reveal how specific design weaknesses influence the likelihood of procedural deviations.

The objectives of this study are threefold: (1) to characterize temporal trends in coupled events as digitalization increases; (2) to quantify the risk amplification effect of interface involvement; (3) to identify the mechanistic pathways driving cognitive misalignment; and (4) Experimental validation through controlled simulator tasks demonstrates the causal mechanism of interface-procedure coupling, elevating the study from descriptive analysis to empirical verification. By integrating factor-level analysis with data-driven modeling, this work establishes an empirical foundation for understanding human-factor vulnerabilities in digital control rooms. The findings highlight that human reliability is increasingly constrained by the semantic and structural qualities of digital interfaces, necessitating integrated design and cognitive-aware monitoring strategies.

This proposed event-based workflow serves as an empirical analytic layer that complements existing third-generation HRA frameworks like IDHEAS-ECA. While IDHEAS-ECA provides a robust top-down cognitive structure, it often lacks site-specific empirical parameters for complex HSI interactions. Our framework functions as a bottom-up validation tool, providing quantitative weights for understanding cognitive failure modes (CFMs). Specifically, we identify interface-procedure coupling as a \textbf{distinct composite PSF}. Unlike traditional frameworks that treat interfaces and procedures as independent PSFs, our approach characterizes their interaction as a \textbf{modifying PSF} that fundamentally alters human error probabilities in digitalized environments.

The remainder of this paper is organized as follows. Section \ref{Literature Review} reviews related literature. Section \ref{Methodology} details the event labeling and modeling methodology. Section \ref{Results and Analysis} presents the quantitative results. Section \ref{Discussion} discusses theoretical and practical implications, and Section \ref{Conclusions and Future Work} concludes with recommendations for enhancing semantic alignment.

\section{Literature Review}\label{Literature Review}
This section reviews prior work on digital HRA and cognitive coupling, and highlights the unresolved empirical challenges that motivate the present event-based analysis.

\subsection{Human Reliability in Digital Environments}
Human reliability analysis (HRA) has undergone a progressive evolution over the past several decades, reflecting advances in both system complexity and cognitive science. The first generation of HRA methods, exemplified by the technique for human error rate prediction (THERP) \cite{kirwan1983technique}, conceptualized human performance largely in terms of discrete task errors within deterministic system models. These approaches provided a foundational taxonomy of human actions and established probabilistic quantification techniques for estimating human error probabilities (HEPs) in safety analyses \cite{nawaz2026understanding}. However, first-generation methods treated human behavior as static and context-independent, offering limited insight into cognitive mechanisms or environmental influences that affect performance.

The second generation of HRA, represented by methods such as the accident sequence evaluation program–human reliability analysis (ASEP) \cite{swain1987accident}, the cognitive reliability and error analysis method (CREAM) \cite{hollnagel1998cognitive}, and the a technique for human event analysis (ATHEANA) \cite{cooper1996technique}, introduced a contextual and cognitive perspective. These frameworks emphasized PSFs \cite{de2011modelling}, including training, workload, team communication, and procedural quality, as determinants of error likelihood. CREAM, in particular, incorporated cognitive control modes to link PSFs with error modes, while ATHEANA sought to represent unsafe acts within the broader context of plant operating conditions \cite{zaitseva2025reliability}. Although these methods significantly advanced the representation of human cognition and situational context, they still relied heavily on qualitative expert judgment and lacked temporal resolution for dynamically changing control environments \cite{briwa2026predicting}.

The advent of digital instrumentation and control (I \& C) systems and computer-based procedures (CBPs) \cite{o2000computer} in modern main control rooms has motivated the development of third-generation HRA frameworks \cite{yu2025aircraft}. These approaches integrate probabilistic reasoning, dynamic modeling, and HSI analytics to address the limitations of earlier generations. The U.S. Nuclear Regulatory Commission’s integrated human event analysis system for event and condition assessment (IDHEAS-ECA) \cite{xiao2025krail} represents a major milestone in this evolution. IDHEAS-ECA explicitly models cognitive mechanisms, such as perception, comprehension, and decision-making, within dynamic task contexts \cite{moafi2026empirical}, providing a bridge between psychological theory and probabilistic safety assessment \cite{yin2025digesting}. Complementary efforts in dynamic HRA employ Bayesian networks \cite{ben2008bayesian}, Markov models \cite{davis2018markov}, and machine-learning architectures to capture time-dependent operator performance and cognitive state transitions in digital control settings.

Overall, although third-generation HRA frameworks such as IDHEAS-ECA have significantly advanced the representation of cognitive processes and contextual dynamics in digital control rooms, they largely remain conceptual or methodologically driven. In particular, existing HRA approaches provide limited empirical grounding for how digital interface characteristics concretely influence procedural execution in real operational settings. Most models still rely on expert judgment or abstract PSF categorizations, with insufficient use of event-level operational data to quantify interface-induced cognitive failures. As a result, there remains a critical gap between cognitively informed HRA theory and empirically validated evidence describing how digital interfaces reshape procedural reliability in practice.

\subsection{Cognitive Coupling and Interface Complexity}
The increasing digitalization of main control rooms has transformed the interaction paradigm between human operators and NPPs from procedural command execution to cognitive coupling with intelligent automation. In this context, cognitive coupling refers to the continuous exchange and alignment of mental models between the operator and the computerized control interface. Effective coupling enables operators to maintain coherent situation awareness and anticipate system responses; conversely, weak or distorted coupling can lead to semantic dissonance, loss of coordination, and ultimately operational errors \cite{zarei2024safety}.

Previous research has demonstrated that as control interfaces evolve toward greater functional integration and automation, interface-induced cognitive workload becomes a dominant factor influencing operator reliability. Ayodeji et al. analyzed the impact of interface complexity on human performance in digitalized nuclear control systems \cite{ayodeji2023cyber}, showing that excessive information density, poor visual hierarchy, and inadequate feedback mechanisms can overload operators’ perceptual and cognitive capacities. Similarly, empirical studies \cite{dickerson2023characterizing} using simulated digital control rooms have found that complex display navigation and inconsistent graphical semantics increase task completion time and error probability, even among experienced operators. These findings underscore that interface usability directly affects the mental workload balance required for safe and timely decision-making.

Recent studies have also explored human–automation coordination failures in digital MCRs, where operators’ mental representations of system status diverge from the system’s actual operational logic. Park et al. \cite{park2022empirical} introduced the concept of cognitive coupling degradation, identifying how mismatched information flows between procedures and digital displays can erode trust calibration, delay fault recognition \cite{li2014fault}, and induce inappropriate interventions. Their results indicate that coupling failures are not isolated ergonomic issues but systemic cognitive breakdowns rooted in inconsistent information semantics. Complementary investigations in other high-reliability domains, such as aerospace and process control, have reported similar phenomena, where automation transparency and feedback delays produce cascading cognitive misalignments between human and machine agents.

Prior studies on cognitive coupling and interface complexity have convincingly demonstrated that semantic inconsistency, interface density, and automation opacity can degrade operator situation awareness and increase cognitive workload \cite{park2022comparisons}. However, the majority of this evidence is derived from laboratory experiments, simulator studies, or qualitative analyses, which limits insight into how frequently such coupling failures occur in actual plant operations and how strongly they amplify procedural deviations. Moreover, existing research often treats interface issues in aggregate, without decomposing their semantic, spatial, and procedural mismatch mechanisms at the event level. These limitations motivate the need for a systematic, event-based empirical analysis capable of quantifying interface–procedure coupling risks and identifying dominant interface mechanisms from real operational data.

Recent studies have further explored the temporal and hierarchical dependencies of PSFs in digital systems. For instance, \cite{liu2025dynamic} utilized DEMATEL-ISM and system dynamics to model the time-varying nature of human error. While they emphasize the longitudinal fluctuation of risk factors, our study provides a complementary focus on the horizontal, semantic coupling between HSI design and procedural logic. Together, these approaches highlight the need for HRA models that integrate both temporal evolution and structural-semantic alignment.

\subsection{Interface–Procedure Interactions and Gaps in Current Research}
Despite growing recognition that human–system interface \cite{o2020human} design plays a decisive role in operator performance within digitalized main control rooms, existing research on interface–procedure interactions remains fragmented and predominantly qualitative. Prior studies have documented how inconsistencies between interface cues and procedural logic can disrupt operators’ mental models, elevate cognitive workload, or induce execution errors \cite{porthin2020effects}. However, these findings are typically derived from controlled simulations, laboratory experiments, or scenario-based cognitive walkthroughs rather than from systematic analyses of real-world operational data. As a result, the empirical basis for understanding the magnitude, frequency, and risk amplification associated with interface-induced procedural deviations remains weak \cite{lundberg2015situation}.

A second limitation concerns the scarcity of event-based quantitative evidence. While several human factors investigations have characterized interface deficiencies, such as ambiguous status indicators \cite{reason1995understanding}, insufficient labeling, and poor layout, few have quantified their direct contribution to procedural deviations or overall operational risk using plant event datasets. Current HRA frameworks acknowledge the possibility of interface-driven impacts but lack empirical parameters describing how often interface problems lead to procedural failures, or under what conditions such coupling effects become safety significant. Consequently, the field lacks validated, data-driven estimates of the causal strength between interface anomalies and procedural errors in operational contexts \cite{lutz2004empirical}.

The third gap lies in the lack of factor-level decomposition of HSI problems. Most prior literature addresses interface issues broadly, referring generically to “complexity,” “ambiguity,” or “usability problems”, without distinguishing their underlying semantic, spatial, feedback, or labeling mechanisms \cite{kim2013detailed}. This limits the capacity to identify which specific interface attributes (e.g., semantic mismatches, mis-hit-prone layouts, inconsistent naming conventions) \cite{roer2019semantic} are most strongly associated with procedural deviations, and which combinations create the conditions for cognitive coupling traps. Without such granularity, design guidelines remain largely generic and insufficiently targeted to the high-impact factors observed in modern digital control rooms.

Collectively, these gaps highlight the need for systematic, event-based, and factor-specific analyses capable of quantifying how interface design interacts with procedural workflows to shape human reliability. Addressing these shortcomings is essential for advancing HRA methodologies and for developing interface and procedural engineering practices that explicitly account for semantic alignment, cognitive integration, and the emergence of interface–procedure coupling traps in next-generation nuclear control rooms \cite{hugo2016method}.

Recent research has increasingly examined human reliability from dynamic and socio-technical perspectives (Table \ref{tab:RESS_review}), with growing attention to cognitive mechanisms, modelling approaches, and operational event analyses. Motivated by these developments, this study investigates interface–procedure interactions in digital control environments using empirical operational data.

\begin{table}[htbp]
\centering
\caption{Recent studies on dynamic human reliability and operator performance in complex socio-technical systems}
\label{tab:RESS_review}
\begin{tabular}{p{3.6cm} p{3.8cm} p{3.2cm} p{5.2cm}}
\toprule
\textbf{Study} & \textbf{Application domain} & \textbf{Methodological focus} & \textbf{Relation to this study} \\
\midrule

 \cite{tan2026dynamic} &
Human–machine collaborative manufacturing &
Dynamic HRA and quality prediction modelling &
Demonstrates the increasing use of dynamic reliability modelling in socio-technical production systems. 
The present study extends this perspective to digital nuclear control rooms by quantifying interface–procedure semantic coupling mechanisms using operational event data. \\

\cite{park2025dynamic} &
Nuclear probabilistic risk assessment &
Dynamic HRA within EMRALD simulation framework &
Provides a simulation-driven dynamic HRA methodology. 
In contrast, the present study offers an event-based empirical quantification of interface-related cognitive risk amplification in real plant operations. \\

\cite{nawaz2026understanding} &
Military aviation maintenance &
PSF–cognitive workload interaction modelling (SEM) &
Highlights the mediating role of cognitive workload in human error formation. 
This study complements such findings by identifying interface–procedure semantic misalignment as a structural PSF mechanism in digitalized control environments. \\

\cite{briwa2026predicting}&
Industrial control room alarm management &
Knowledge-based Bayesian network reliability prediction &
Demonstrates operator reliability modelling through cognitive-task decomposition. 
The present work advances this line by incorporating semantic interface factors derived from real incident datasets. \\

\cite{yin2025digesting} &
Nuclear power plant commissioning events &
Incident-based HRA coding methodology &
Provides an integrated framework for analysing human-related incidents in nuclear operations. 
This study builds upon this event-oriented paradigm by introducing a factor-level decomposition of interface-procedure coupling risks. \\

\cite{yu2025aircraft} &
Aircraft maintenance scheduling &
Human-error-aware optimisation modelling &
Shows the operational impact of human reliability on system performance. 
The present study contributes complementary insights by focusing on cognitive integration failures in digital control interfaces. \\

\cite{zaitseva2025reliability}&
Healthcare reliability engineering &
AI-enabled reliability assessment in complex systems &
Illustrates the broad expansion of reliability engineering methods across human-centric domains. 
This study situates digital nuclear HSI as a similarly emerging application requiring integrated cognitive-risk modelling. \\

\cite{moafi2026empirical} &
Human–robot interaction systems &
Reliability enhancement via advanced control strategies &
Emphasizes the role of dynamic system modelling in ensuring safety in human-interactive technologies. 
The present study parallels this direction by analysing reliability challenges in digitalized operator–interface ecosystems. \\

\bottomrule
\end{tabular}
\end{table}

\section{Methodology}\label{Methodology}

\subsection{Overview of the Framework}\label{Overview of the Framework}
The overall methodological framework integrates structured expert labeling, interface-factor decomposition, statistical inference, and interpretable machine-learning techniques to systematically quantify the emergence and properties of interface–procedure coupling risks. As illustrated in Figure \ref{framework}, the analytical pipeline begins with event collection, cleaning, and semantic harmonization to ensure consistency across multi-year operational records. Cleaned events are then subjected to a standardized labeling schema capturing procedural deviations (PROC), interface-related issues (UI), and coupled interface–procedure failures (COUPLED). For all UI-positive cases, a secondary decomposition identifies the specific interface design factors underlying the deviation, enabling a more granular representation of interface-induced vulnerabilities.

\begin{center}
\includegraphics[width=0.6\textwidth]{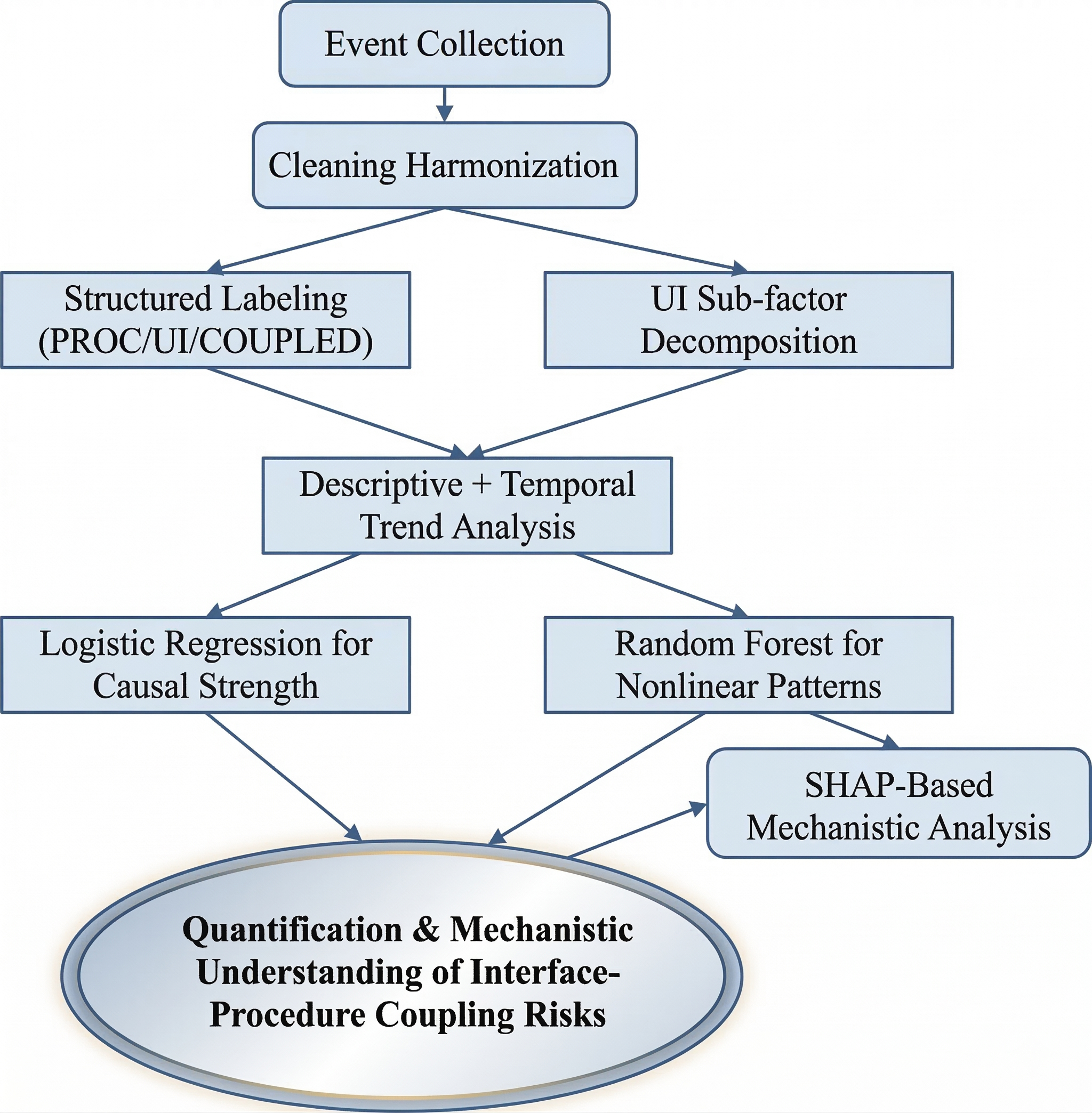}
\captionof{figure}{Integrated analytical framework for quantifying interface–procedure coupling risks, combining structured event labeling, interface-factor decomposition, statistical modeling, and interpretable machine learning.}\label{framework}
\end{center}

Subsequent stages of the analysis integrate descriptive statistics and temporal trend assessment to characterize shifts in human-factor risk mechanisms accompanying increasing digitalization. Logistic regression is applied as a parsimonious, hypothesis-driven approach to estimate the direction and approximate magnitude of the association between interface involvement and procedural deviations, while a Random Forest classifier provides a complementary, data-driven perspective by capturing nonlinear dependencies and interaction patterns among interface design factors and contextual variables under limited sample size conditions. SHAP-based interpretability techniques are further employed to support event-level mechanistic interpretation of how specific interface features contribute to procedural failure likelihood. Collectively, this multi-layered analytical strategy is intended not to establish population-level causal estimates, but to provide empirical, event-based evidence of interface–procedure coupling mechanisms grounded in real operational data, and to assess their relative tendencies and internal consistency across observed cases, thereby supporting both quantitative association analysis and mechanistic understanding of coupling-related risk in digital control environments.

\subsection{Dataset and Scope} \label{Dataset and Scope}
This study draws on a corpus of formally classified A-class and B-class nuclear operational event reports collected from multiple commercial NPPs in China between 2021 and 2025. These reports are standardized, regulator-reviewed records within the national operational experience feedback system and reflect real operational events rather than simulated scenarios or questionnaire-based observations. A-class events involve occurrences with significant implications for plant safety, equipment integrity, radiation protection, or operational availability, whereas B-class events include reportable deviations, equipment anomalies, maintenance-related issues, and other events with diagnostic or trend-analysis value. Together, these event classes span a wide range of human–system interaction contexts, including routine operations, maintenance, testing, and digital MCR activities.

Each report contains narrative descriptions, root-cause analyses, corrective actions, and relevant operational context, enabling structured identification of procedural deviations, interface-related deficiencies, and interface–procedure coupling mechanisms. All data were fully de-identified and analyzed solely for research purposes. The scope of this study is intentionally focused on identifying recurrent human-factor patterns and examining how interface semantics influence procedural reliability in modernized digital control environments.

It is acknowledged that analyses based on formal operational event reports are subject to reporting bias and incompleteness. While A- and B-class reports reliably capture safety-significant deviations, they may underrepresent interface-related near-misses, transient cognitive difficulties, or short-lived semantic inconsistencies that are resolved by operators without escalating into reportable events. Accordingly, the dataset reflects the observable and consequential portion of interface–procedure coupling risk, and the reported coupling frequencies should be interpreted as conservative lower-bound estimates rather than exhaustive representations. Although organizational and training factors are not modeled explicitly as independent predictors, their potential influence is partially reflected in event narratives and is discussed qualitatively in Section \ref{Discussion}.

Finally, it should be emphasized that high-quality operational event data in nuclear power plants are inherently scarce, particularly for formally classified A- and B-class events. The present dataset covers all available A/B-class events associated with digitalized control room operations during the 2021–2025 period within the participating plants, and thus constitutes an operationally complete registry of regulator-reportable safety-significant events within the defined study boundary rather than a statistical sample. As such, conventional assumptions of large-sample representativeness and statistical generalization are not directly applicable.

In this study, the analytical population is explicitly defined as regulator-recognized safety-significant human-factor events occurring in digitalized control-room contexts within the observation period. Accordingly, the dataset should not be interpreted as a statistical sample of all human–system interaction difficulties, but rather as a bounded empirical population representing events that have exceeded multiple operational defense layers and triggered formal reporting mechanisms. This risk-threshold-based definition allows the study to focus on the mechanisms underlying operationally consequential cognitive failures, which constitute the most safety-relevant subset of interface–procedure coupling phenomena.

\subsection{Event Labeling Framework} \label{Event Labeling Framework}
To enable systematic analysis of HSI failures, all events in the dataset were annotated using a structured binary labeling framework. The framework distinguishes between \textit{procedure-related} deviations, \textit{interface-related} issues, and \textit{coupled} events in which interface deficiencies trigger or amplify procedural deviations. Labels were assigned based on explicit evidence extracted from the event narratives, causal analyses, and corrective action records.

\textbf{Procedure-related deviations.} An event was labeled as procedure-related if any of the following conditions were met: (i) failure to follow prescribed procedures, including step omission, step skipping, incomplete execution, or explicit procedural violations; (ii) deficiencies in the procedure itself, such as missing steps, ambiguous or outdated content, or incorrectly specified actions; (iii) unclear or insufficiently specified verification or confirmation steps; (iv) absence of required procedural warnings or risk cues that contributed to operator misjudgment or execution errors.

\textbf{Interface-related issues.} An event received an interface-related label if it included evidence of human–machine interface deficiencies, including: (i) unclear, inconsistent, or missing labels, indicators, or legends; (ii) confusing spatial layout, indistinguishable panel sections, or selection errors caused by similar cabinet or interval appearance; (iii) incorrect or inconsistent state indications, such as valve position displays or pointer orientations not aligned with actual equipment status; (iv) elevated risk of inadvertent actuation due to dense control layouts, poor affordance, or insufficient visual salience of control states.

\textbf{Coupled interface–procedure deviations.} Events that simultaneously exhibited both procedural and interface-related characteristics were labeled as \textit{coupled}. These cases represent typical human-factor traps in modernized control environments, where interface deficiencies induce, misguide, or magnify deviations from procedural logic. Coupled events were retained as an explicit category given their relevance to cognitive–semantic mismatches in digital MCRs and their disproportionately high contribution to operational risk.

In cases where interface-related issues appeared in event narratives but were not explicitly identified as causal factors in the root-cause analysis, a conservative labeling strategy was applied: events were classified as interface-related only when explicit narrative evidence indicated that interface characteristics plausibly influenced operator perception, decision-making, or action selection during event progression; otherwise, such events were coded as procedure-related only, even if interface deficiencies co-existed in the background. Interface–procedure coupled events (COUPLED = 1) were assigned strictly when both a procedural deviation or procedural inadequacy was identified and interface characteristics were shown to induce, misguide, or amplify the deviation through mechanisms such as semantic inconsistency, spatial confusion, or feedback ambiguity, while events in which interface issues were incidental, latent, or unrelated to the procedural outcome were intentionally excluded.

All event labels were assigned according to predefined coding rules and independently reviewed by two analysts. \textbf{To ensure the reliability of the qualitative coding, the initial independent agreement level was assessed, reaching over 85\% across all primary categories (PROC, UI, and COUPLED).} Borderline cases or discrepancies—\textbf{which typically arose from complex narratives where interface issues were latent rather than explicit}—were resolved through a structured consensus discussion. This consensus-based approach was preferred over simple statistical averaging as it allowed for a deeper, expert-driven reconciliation of technical event details, ensuring that the labeling preserved sensitivity to subtle interface-induced cognitive mechanisms. This structured and conservative labeling framework enables a quantitative assessment of the relative contributions of procedural quality, interface design, and their interaction effects as a composite PSF in real operational events.

For all identified UI-positive events ($n=22$), a secondary decomposition was performed using a hierarchical decision tree to categorize specific failure mechanisms. The classification followed a rigorous three-step triangulation protocol to ensure diagnostic validity (Figure \ref{fig:ui_decision_tree}). First, the primary cognitive disruption was identified by distinguishing whether the error stemmed from spatial orientation ($UI_{Layout}$), interpretation of information ($UI_{Semantic}$), procedural inconsistency ($UI_{Mismatch}$), or component identification ($UI_{Label}$). Second, events were mapped against specific diagnostic criteria; for instance, $UI_{Layout}$ was assigned in cases of spatial confusion or proximity-induced errors (e.g., mis-hitting adjacent buttons), while $UI_{Semantic}$ was reserved for ambiguous state indications or misleading symbolism. Finally, assignments were cross-validated by analyzing the corresponding corrective actions—such as verifying that a physical barrier installation confirmed a layout issue, or a logic revision confirmed a semantic one. This integrative approach, linking event narratives, causal evidence, and remediation strategies, ensured that the UI factor classifications were grounded in empirical data and internally consistent.

\begin{center}
\centering
\includegraphics[width=1.0\textwidth]{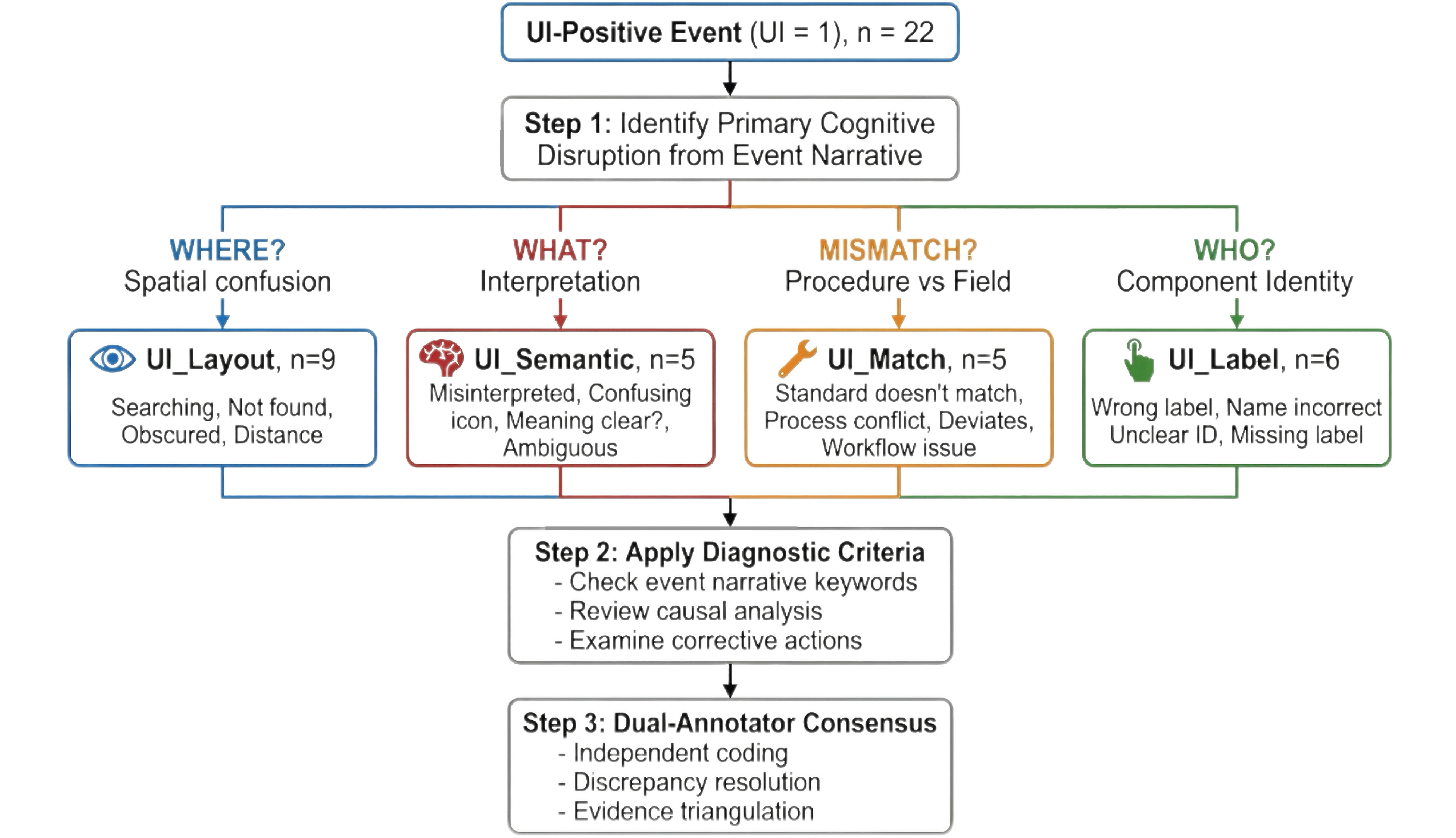}
\captionof{figure}{Hierarchical decision tree for systematic extraction of UI sub-factors from event narratives.}\label{fig:ui_decision_tree}
\end{center}

To ensure the reliability and reproducibility of the event labeling process, a structured dual-annotator protocol was implemented. Two analysts with expertise in nuclear human factors independently reviewed all 59 events using a standardized coding manual. Each analyst conducted independent coding for the PROC, UI, and COUPLED categories based on predefined diagnostic criteria

Inter-rater agreement was subsequently assessed using Cohen’s kappa ($\kappa$), demonstrating high consistency across all categories (PROC: $\kappa$ = 0.89, UI: $\kappa$ = 0.82, COUPLED: $\kappa$ = 0.85), indicating substantial to almost perfect agreement. For the eight discrepant cases (13.6\%), a structured consensus process was conducted. Analysts jointly reviewed event narratives, root-cause analyses, and documented corrective actions, and final labels were determined through evidence-based discussion rather than majority voting.

Typical discrepancies arose from three patterns. First, in several cases interface deficiencies were mentioned in event descriptions but were not clearly identified as causal contributors. These events were coded as UI = 1 only when explicit evidence demonstrated that interface characteristics influenced operator perception or action selection. Second, disagreements occasionally occurred in distinguishing procedural inadequacy from operator non-compliance. In such cases, PROC = 1 was assigned when either procedural content was insufficient or operators deviated from prescribed steps. Third, ambiguity was observed in determining whether procedural and interface issues constituted coupled failures. COUPLED = 1 was assigned only when interface characteristics were shown to directly induce or amplify the procedural deviation.

Overall, this consensus-based labeling strategy enabled the identification of subtle interface-induced cognitive mechanisms while maintaining a high level of inter-rater reliability, thereby enhancing the methodological robustness of the dataset.

\subsection{Decomposition of Interface Design Factors} \label{Decomposition of Interface Design Factors}
For all events classified as interface-related, a secondary decomposition was conducted to identify the specific interface mechanisms contributing to the observed deviations. This analysis yielded four non-exclusive categories that capture distinct dimensions of HSI design in digital MCRs. The first category, $UI_{Layout}$, encompasses issues related to spatial arrangement, mis-hit actions, navigation difficulties, and suboptimal operability zones. The second category, $UI_{Semantic}$, reflects ambiguities or misleading aspects of interface semantics, including unclear status indicators, reversed or counterintuitive valve direction displays, and insufficiently interpretable graphical elements. The third category, $UI_{Mismatch}$, denotes semantic inconsistencies between procedural expectations, engineering drawings, and actual interface representations, such as discrepancies between indicator logic and the procedural steps an operator is expected to follow. The final category, $UI_{Label}$, captures deficiencies in labeling and nomenclature, including missing labels, inconsistent naming conventions, and non-salient identifiers that hinder accurate recognition.

Each category was coded as present when the corresponding mechanism contributed to the event, allowing multiple mechanisms to be simultaneously attributed to a single event. This structured decomposition is consistent with established taxonomies in human–system interface design and supports a more granular examination of how interface features influence operator performance in digitalized control environments.

The selection of the four interface design factors: layout, semantic, mismatch, and labeling, was guided by both established HSI taxonomies and the empirical characteristics of the event dataset. These factors correspond to the core stages of operator information processing, namely perception (layout and labeling), interpretation (semantic), and cognitive integration between procedural intent and interface representation (mismatch). Alternative interface dimensions discussed in the literature, such as feedback timing or automation transparency, were considered during the preliminary coding phase. However, these aspects were found to be either implicitly embedded within semantic and mismatch issues in the analyzed event reports or insufficiently documented in a consistent manner to support reliable event-level labeling. Therefore, the adopted four-factor decomposition represents a balance between theoretical completeness and empirical traceability, ensuring that each category is both cognitively meaningful and operationally observable in real-world event data.

\subsection{Statistical Analysis and Hypothesis Testing} \label{Statistical Analysis and Hypothesis Testing}
\subsubsection{Descriptive statistics} 
Following the structured labeling process, descriptive statistical analyses were performed to evaluate the distribution and characteristics of PROC, UI, and COUPLED events. Annual frequencies, co-occurrence patterns, and proportional contributions of each category were computed to assess temporal changes in event mechanisms over the 2021–2025 study period. Visualization tools, including stacked bar charts and temporal heatmaps, were used to depict evolving risk patterns and to identify trends associated with increasing interface complexity and digitalization.

\subsubsection{Logistic regression} 
To quantify the effect of interface involvement on the likelihood of procedural deviation, a binary logistic regression model was fitted with PROC as the dependent variable and UI as the sole predictor:

\begin{equation}
    logit(Pr(PROC=1))=\beta _{0}+\beta _{1}\cdot UI
\end{equation}

The model yielded parameter estimates from which odds ratios (OR), 95\% confidence intervals, and corresponding p-values were derived. Diagnostic checks, including examination of residuals, leverage, and indicators of separation, were performed to assess model stability given the modest sample size. Notably, certain UI sub-factors exhibited quasi-complete separation (e.g., all cases involving $UI_{Mismatch}$ were also procedure-related), which limited the feasibility of fitting a stable multivariable logistic regression model including all interface factors simultaneously. To address this limitation, logistic regression was supplemented with machine-learning–based analyses, which are better suited to capturing nonlinear effects and interactions under conditions of small sample size and partial feature separation.

\subsection{Machine Learning Classification Analysis} \label{Machine Learning Classification Analysis}
\subsubsection{Model Design}
To complement the hypothesis-driven logistic regression and to explore potential nonlinear interactions and explanatory patterns among interface-related factors, a Random Forest (RF) classifier was employed. Given the modest sample size, the RF model is not intended as a predictive tool, but rather as an exploratory and interpretive method to identify relative factor importance and interaction tendencies that may not be captured by parametric models. The model incorporated a set of binary predictors representing the presence of interface involvement (UI) and four specific UI design sub-factors, layout/navigation issues ($UI_{Layout}$), semantic ambiguity ($UI_{Semantic}$), procedure–interface mismatches ($UI_{Mismatch}$), and labeling deficiencies ($UI_{Label}$). In addition, a standardized Year variable was included to capture potential temporal trends associated with digitalization and interface modifications across the 2021–2025 period. The dependent variable was the binary indicator PROC, which denoted whether an event involved a procedure-related deviation (PROC = 1). This modeling structure allowed the RF classifier to detect complex, high-order patterns that may not be adequately captured by conventional parametric approaches.

\subsubsection{Training and Validation}
Model training employed a five-fold stratified cross-validation procedure to address class imbalance and to ensure robust generalization across varying event compositions. For each fold, classification accuracy, fold-wise variance, and out-of-bag (OOB) error were computed to evaluate model stability and performance. Feature importance values were derived using the mean impurity decrease (Gini importance), providing a global measure of each predictor’s contribution to the RF decision process. This validation strategy ensured that the classifier’s performance metrics reflected both predictive reliability and resistance to overfitting, despite the relatively small sample size.

Accordingly, model performance metrics (e.g., classification accuracy) are reported to demonstrate internal consistency rather than predictive reliability, and SHAP-based interpretation is used to support mechanistic understanding of interface–procedure coupling rather than outcome prediction.

\subsubsection{SHAP-Based Interpretation}
To complement the global importance metrics and enhance interpretability, SHapley Additive exPlanations (SHAP) \cite{antwarg2021explaining} analysis was employed to quantify both local and global feature contributions. Two forms of SHAP visualizations were generated: (a) a SHAP summary plot illustrating the distribution and magnitude of each predictor’s marginal impact on the model output across all events, and (b) a SHAP force plot for a representative coupled event, highlighting how specific UI-related factors shifted the predicted probability of a procedural deviation. This interpretable machine-learning approach provides mechanistic insight into how interface semantics, design inconsistencies, and contextual factors jointly influence the emergence of procedure-related failures at an instance-specific level.

\subsection{ Experimental Validation}
To empirically validate the interface–procedure coupling mechanisms identified from historical operational events, a controlled simulator-based experiment focusing on parameter-verification tasks was conducted (Figure \ref{experiment}). The experiment comprised five task scenarios, each including six to seven cross-system verification tasks, yielding a total of 518 verification operations. Representative coupling traps derived from the analysis of the 59 operational events were deliberately embedded into the task design. These traps included semantic ambiguities between similarly named subsystems (e.g., Nuclear Island System versus Nuclear Island Auxiliary System), labeling similarity among parameter identifiers (e.g., 2LBA10CP801A, 2LBA10CP801B, and 2LBA10CP801C), and layout dispersion of relevant parameters across multiple display windows such as LAB DW001, KBE DW101, and PCB DW001.

\begin{center}
\centering
\includegraphics[width=1.0\textwidth]{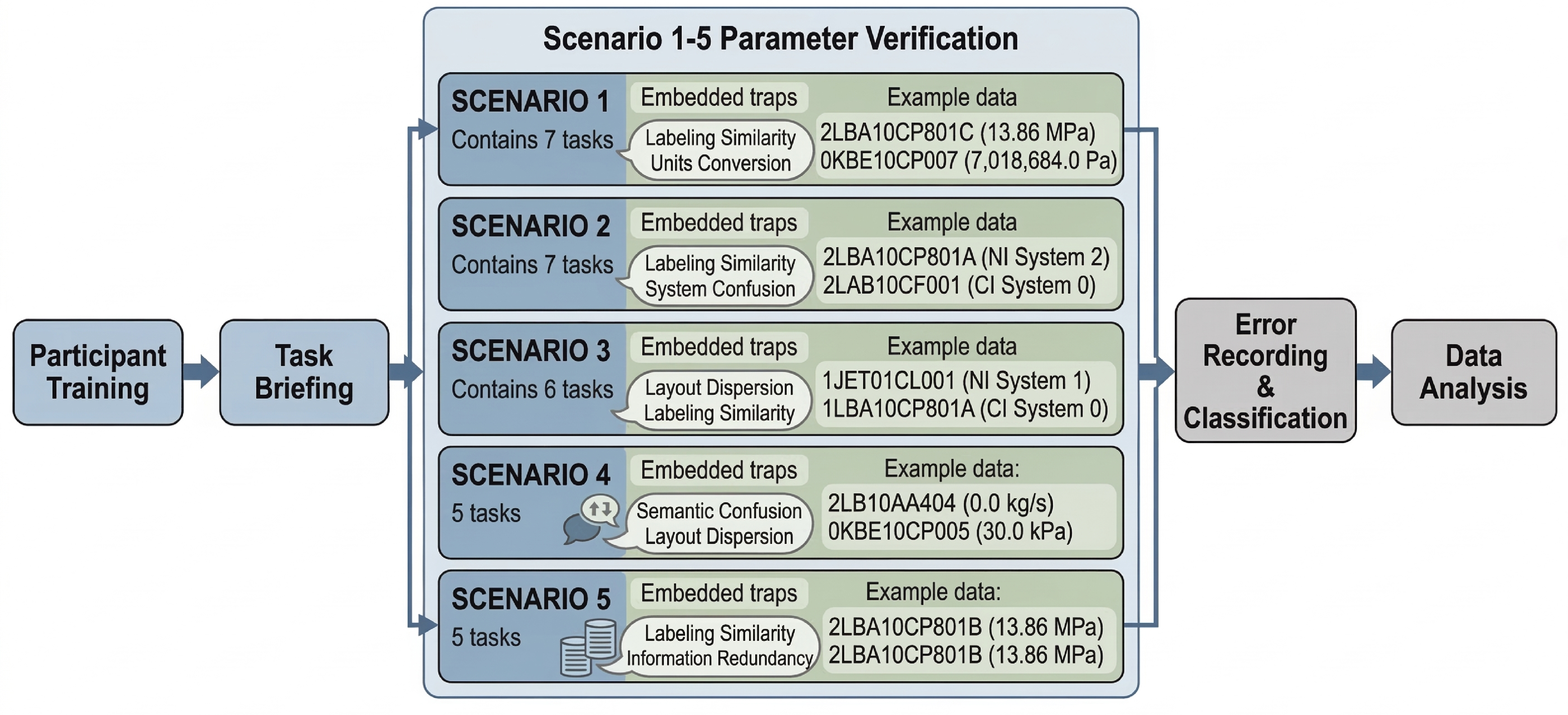}
\captionof{figure}{Experimental procedure for validating interface-procedure coupling traps. Five scenarios were designedwith embedded coupling traps including semantic confusion, layout dispersion, and labeling similarity.}\label{experiment}
\end{center}

Six graduate students with prior training in nuclear power plant operation participated in the experiment. Operator interactions were continuously recorded using eye-tracking and screen-capture techniques to enable detailed behavioral analysis. Observed performance deviations were subsequently classified into five categories: confusion errors, click errors, recording errors, recognition errors, and hesitation or task abandonment.

\section{Results and Analysis}\label{Results and Analysis}

This section focuses on descriptive statistics and observed temporal patterns. Interpretations regarding underlying mechanisms and potential causal explanations are intentionally deferred to Section \ref{Discussion}.

\subsection{Event Labeling and Descriptive Statistics}\label{Event Labeling and Descriptive Statistics}
This section presents the structured labeling results for all A-class and B-class events from 2021–2025, followed by descriptive statistics quantifying the prevalence of procedure-related deviations, interface-related deficiencies, and coupled interface–procedure traps. A total of nine A-class events were analyzed, each reflecting high-significance operational deviations with potential safety or availability implications. These events were systematically annotated along three dimensions: procedure-related, interface-related, and coupled (interface-induced procedural deviation). The labeling results show that 66.7\% of A-class events involved procedural deviations, 33.3\% involved interface deficiencies, and 22.2\% exhibited coupled characteristics. Notably, the two coupled A-class events originated from MCR activities, indicating that high-impact events in contemporary digital control environments frequently arise from simultaneous procedural and interface failures rather than from procedural non-compliance alone.

In addition to the A-class sample, forty-seven B-class events from 2021 to 2025 were fully cleaned and reclassified using the same labeling framework. Clear temporal patterns emerged across the five-year period. In 2025 (n = 11), procedure-related and interface-related events occurred with equal frequency (both n = 6), with four cases (67\% of interface-related) classified as coupled, underscoring the strong interaction between UI deficiencies and procedural execution. The 2024 dataset (n = 12) showed a similar pattern, with nine procedure-related events, five interface-related events, and five coupled cases, indicating stable co-occurrence between UI inconsistencies and procedural deviations. In 2023 (n = 11), four interface-related and four coupled events were documented, many of which involved misinterpretation of labels or insufficient visual distinctiveness leading to step omissions or incorrect verification actions. The 2022 sample (n = 13) was dominated by procedural deviations (n = 11) but also demonstrated five interface-related and five coupled events, driven primarily by naming inconsistencies and ambiguous indication of valve positions. By contrast, the 2021 sample (n = 3) contained exclusively procedure-related deficiencies with no interface-related issues, reflecting the relative absence of digital-interface complexity during that period.

Aggregated across all B-class events (n = 47), procedure-related deviations accounted for 80.9\% of cases, interface-related issues for 42.6\%, and coupled events for 38.3\%. A particularly salient finding is that 90\% of interface-related events simultaneously exhibited procedural deviations. This indicates that interface deficiencies almost never occur in isolation; instead, they tend to trigger, shape, or amplify procedural errors, thereby forming system-level human-factor traps. Such structural coupling is a defining characteristic of operational risk in digitalized main control rooms, where UI semantics and procedural logic increasingly interact to influence operator cognition and task performance.

Temporal trend analysis (Figure \ref{event_analysis} a) underscores these observations. Interface-related events increased markedly from 2021 through 2025. This increase temporally coincides with the progressive digitalization of main control room interfaces during the same period. Coupled events remained consistently high (approximately 35–40\% annually), indicating persistent and systemic interaction between interface design and procedure execution. Strikingly, in 2025, the number of interface-related events equaled that of procedure-related events for the first time, indicating a notable shift in the observed distribution of event mechanisms. Taken together, the descriptive statistics indicate a strong co-occurrence between interface-related issues and procedural deviations in the analyzed events.

The collected A-class and B-class reports are presented in Appendix \ref{A-Class Event Reports} and Appendix \ref{B-Class Event Reports (2025–2021)}, respectively. The detailed contents are provided in Tables \ref{tab:A_class_events}, \ref{tab:B_class_2025}, \ref{tab:B_class_2024}, \ref{tab:B_class_2023}, \ref{tab:B_class_2022}, and \ref{tab:B_class_2021}.

\begin{center}
\includegraphics[width=1.0\textwidth]{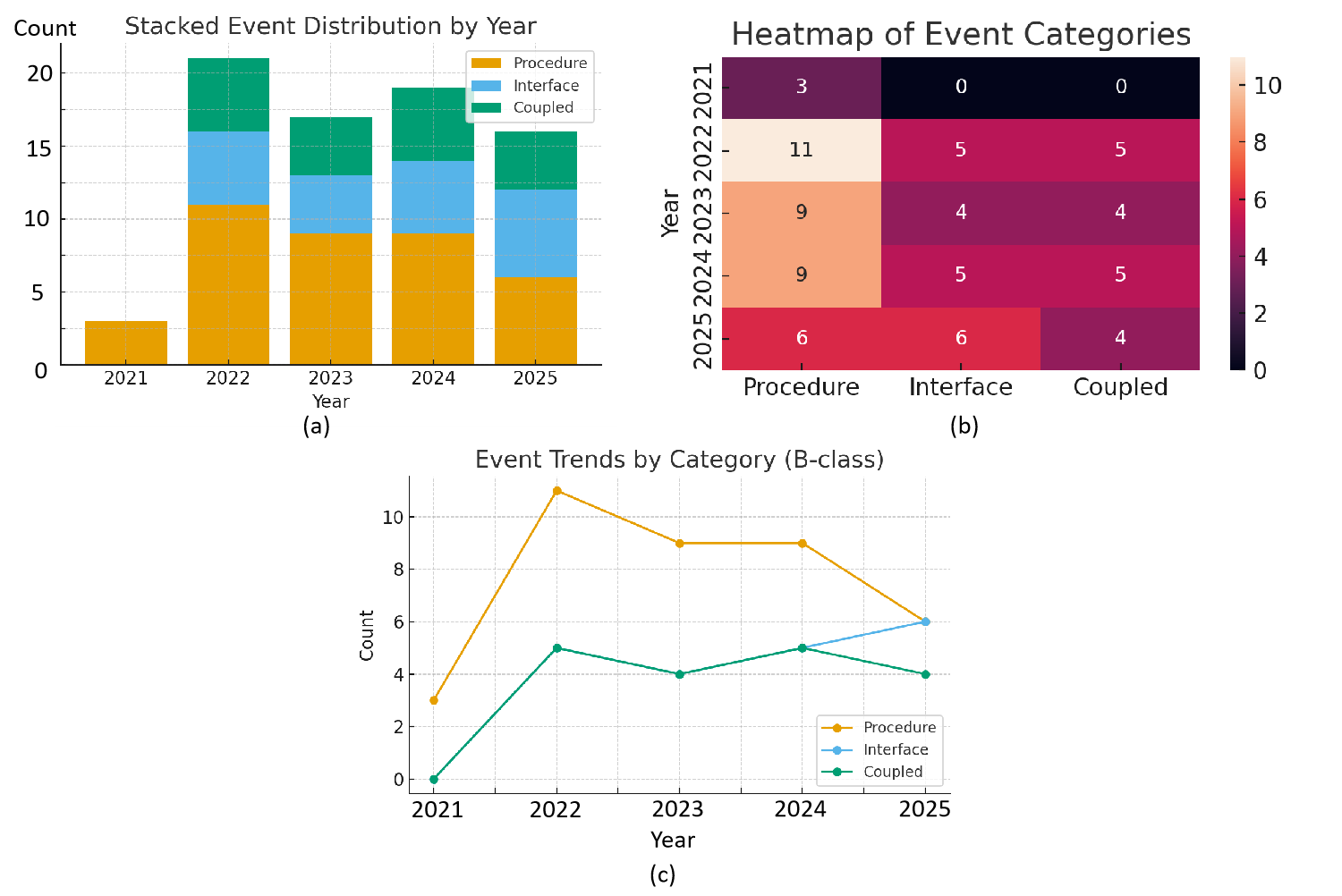}
\captionof{figure}{Annual Distribution, Frequency Heatmap, and Temporal Trends of Procedure, Interface, and Coupled Events (2021–2025).}\label{event_analysis}
\end{center}

\subsection{Event Distribution and Trends}\label{Event Distribution and Trends}
Human-factor risks in nuclear power plant main control rooms are undergoing a clear structural shift. Our multi-year analysis (Figure \ref{Event_trends} (a)) shows that event mechanisms are transitioning from traditional procedure deviations toward interface-related and interface–procedure coupling traps. Interface factors, once marginal, now account for 35–50\% of annual A/B-class events, and in 2025, they surpass procedure-related factors for the first time. More importantly, coupled failures, where both procedural guidance and interface semantics break down simultaneously, remain persistently high at 35–40\%, indicating that digitalized control environments inherently amplify the interaction between procedural intent and interface design.

The rise of coupling traps is particularly pronounced within digital main control rooms, where their share increases from 20\% in 2022 to 75\% in 2025 (Figure \ref{Event_trends} (b)). This trend suggests that cognitive misalignment between interface elements and procedural actions has become a primary safety challenge. As interface density, semantic layering, and interaction pathways grow more complex, operators face a greater risk of misinterpreting displays, selecting incorrect channels, or executing steps out of sequence, even when procedures themselves are well-structured.

Overall, these findings provide quantitative evidence that the dominant mode of human-factor risk in modern NPPs is no longer single-source procedural deviation but multi-source cognitive coupling. Ensuring safety in next-generation digital control rooms therefore requires coordinated improvements in interface design, procedural engineering, and real-time cognitive evaluation tools capable of identifying semantic misalignment before it propagates into operational error.

\begin{center}
\centering
\includegraphics[width=1.0\textwidth]{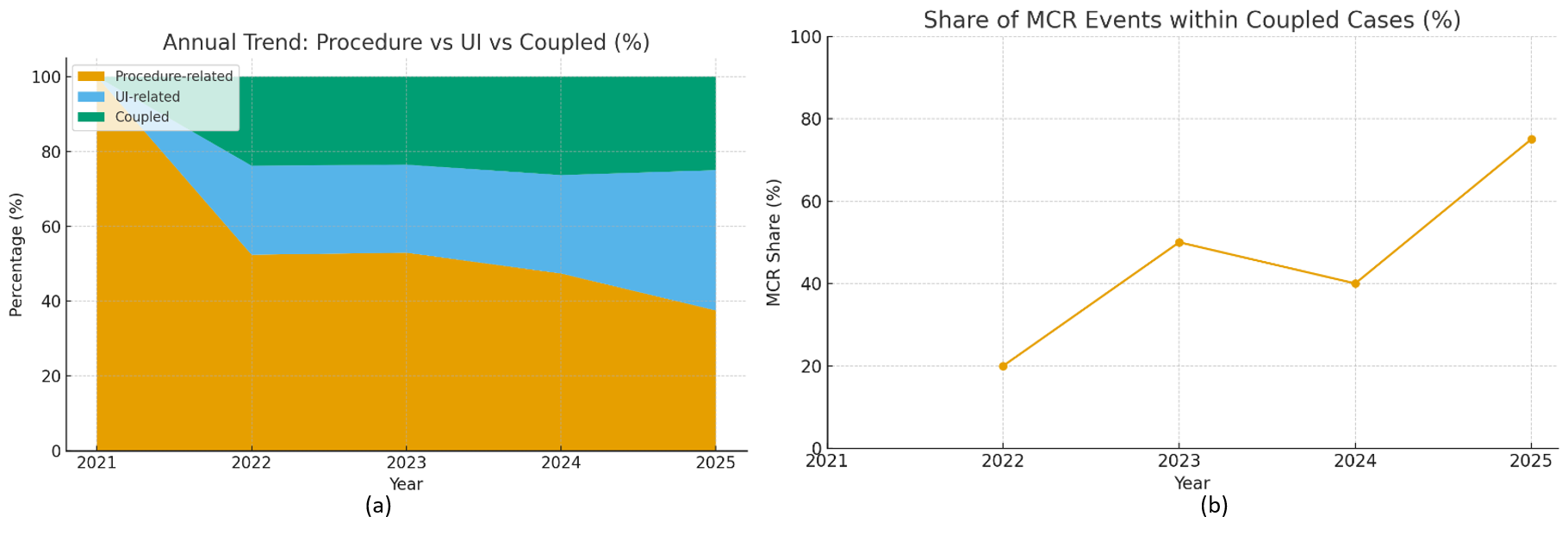}
\captionof{figure}{Annual Distribution, Frequency Heatmap, and Temporal Trends of Procedure, Interface, and Coupled Events (2021–2025).}\label{Event_trends}
\end{center}

Additionally, the proportion of interface–procedure coupling events occurring within the MCR shows a clear upward trend across the study period. Specifically, MCR-attributed coupled events increased from 20\% in 2022 to 50\% in 2023, remained elevated at 40\% in 2024, and rose markedly to 67\% in 2025. This escalation indicates that coupled failures are progressively concentrating within increasingly digitalized control-room environments. The trend suggests that insufficient semantic support in interface design is allowing discrepancies between interface cues and procedural logic to intensify into operator deviations during MCR operations. Accordingly, next-generation human–factors tools for digital control rooms must prioritize the alignment of interface semantics with procedural intent to mitigate the growing prevalence of coupling-related risks.

These temporal patterns are reported descriptively and should not be interpreted as evidence of a causal effect of digitalization, given the observational nature of the event data.

\subsection{Severity Correlation}\label{Severity Correlation}

Within the limited set of high-severity cases, interface-driven and interface–procedure coupling mechanisms appear more frequently than in lower-severity events. High-severity cases frequently involve irreversible operational consequences, such as unintended reactor trips or incorrect protection-parameter switching, typically triggered by a combination of UI-induced misinterpretation and procedural ambiguity. Other recurrent patterns include button or valve-state misjudgment, often in contexts lacking explicit confirmation steps, indicating that interface-evoked procedural deviations play a central role. Collectively, these findings suggest a strong \textbf{association} between UI-related factors and high CR levels, with interface traps exhibiting a distinct \textbf{risk amplification pattern} compared to standalone procedural violations. These proportions highlight that the primary \textbf{factors associated with} high-risk events are not simply procedural non-compliance, but rather semantic discontinuities between procedural intent and interface representation.

A decomposition of causal sub-factors further clarifies the underlying structure of these failures. Approximately 45\% of events involve unclear or insufficient procedural content (e.g., missing verification steps, ambiguous instructions). Interface–procedure misalignment, such as discrepancies between valve status, indicator directions, and procedural logic, accounts for 38\%. Deficiencies in labeling and layout, leading to inadvertent activation or navigation errors, contribute 32\%, while insufficient operator training or limited experience accounts for 21\%. These proportions highlight that the primary drivers of high-risk events are not simply procedural non-compliance, but rather semantic discontinuities between procedural intent and interface representation.

Overall, the results suggest that, among the analyzed cases, semantic gaps between procedural intent and interface representation are a recurring contributor to high-severity outcomes, although definitive attribution is constrained by the sample size. Addressing this gap constitutes the central challenge that emerging intelligent HRA and cognitive-aware control-room tools must resolve. It should be noted that the number of high-consequence (CR) events in the dataset is relatively limited. Accordingly, the observed associations between interface-related factors and event severity should be interpreted with caution. The reported patterns indicate descriptive correlations rather than statistically robust or generalizable severity predictors.

\subsection{Statistical Tests}\label{Statistical Tests}

A total of 59 A-class and B-class events (2021–2025) were included in the analysis after cleaning and structured recoding. Among these events, 46 (77.97\%) were classified as procedure-related (PROC = 1), 22 (37.29\%) involved interface-related issues (UI = 1), and 19 (32.20\%) were labeled as coupled events where interface problems induced or directly contributed to procedural deviations (COUPLED = 1). Table \ref{tab:desc_stats} summarizes the joint distribution of procedure-related and interface-related codings:

\captionof{table}{Descriptive statistics of A- and B-class event dataset (2021--2025).}\label{tab:desc_stats}
\begin{center}
\begin{tabular}{lcc}
\hline
\textbf{Statistic} & \textbf{Count} & \textbf{Percentage} \\
\midrule
Total events & 59 & 100\% \\
Procedure-related (PROC = 1) & 46 & 77.97\% \\
Interface-related (UI = 1) & 22 & 37.29\% \\
Coupled events (COUPLED = 1) & 19 & 32.20\% \\
\hline
\end{tabular}
\end{center}

In the absence of interface issues, 72.97\% (27/37) of events were still procedure-related, reflecting the dominant role of procedural weaknesses in the dataset. However, when interface problems were present, the proportion of procedure-related events increased to 86.36\% (19/22), suggesting a potential amplifying effect of UI issues on procedural deviations. Coupled events (PROC = 1 and UI = 1 coded as COUPLED = 1) were observed across all years except 2021, with concentrations in 2022–2025, consistent with the growing prevalence and complexity of digital interfaces in recent plant modifications and operational activities.

\subsection{Association between interface involvement and procedure-related deviations}

To quantify the association between interface problems and procedural deviations, a binary logistic regression was fitted with PROC (1 = procedure-related, 0 = not procedure-related) as the dependent variable and UI (1 = interface-related, 0 = not interface-related) as the sole predictor:

\begin{equation}
    logit(Pr(PROC=1))=\beta _{0}+\beta _{1}\cdot UI
\end{equation}

The model converged and yielded a positive, though not statistically significant, coefficient for UI. The estimated odds ratio (OR = 2.35) indicates a positive association between interface involvement and the likelihood of procedure-related deviations. Although this link does not reach conventional levels of statistical significance (p = 0.24), the direction and magnitude of the estimate suggest a potentially meaningful trend rather than a definitive causal effect. Given the limited sample size and the presence of quasi-complete separation, the regression results should be interpreted as exploratory evidence supporting the plausibility of a risk amplification mechanism, rather than as a precise quantification of causal strength. Importantly, this trend is consistent with the descriptive statistics and the machine-learning analyses presented in Sections \ref{Event Labeling and Descriptive Statistics} and \ref{Random Forest Classification and Feature Importance}, which independently indicate that interface-related deficiencies tend to co-occur with procedural deviations. Figure \ref{figure_mix} (a) visualizes the point estimate and 95\% confidence interval on a logarithmic scale.

\begin{center}
\centering
\includegraphics[width=1.0\textwidth]{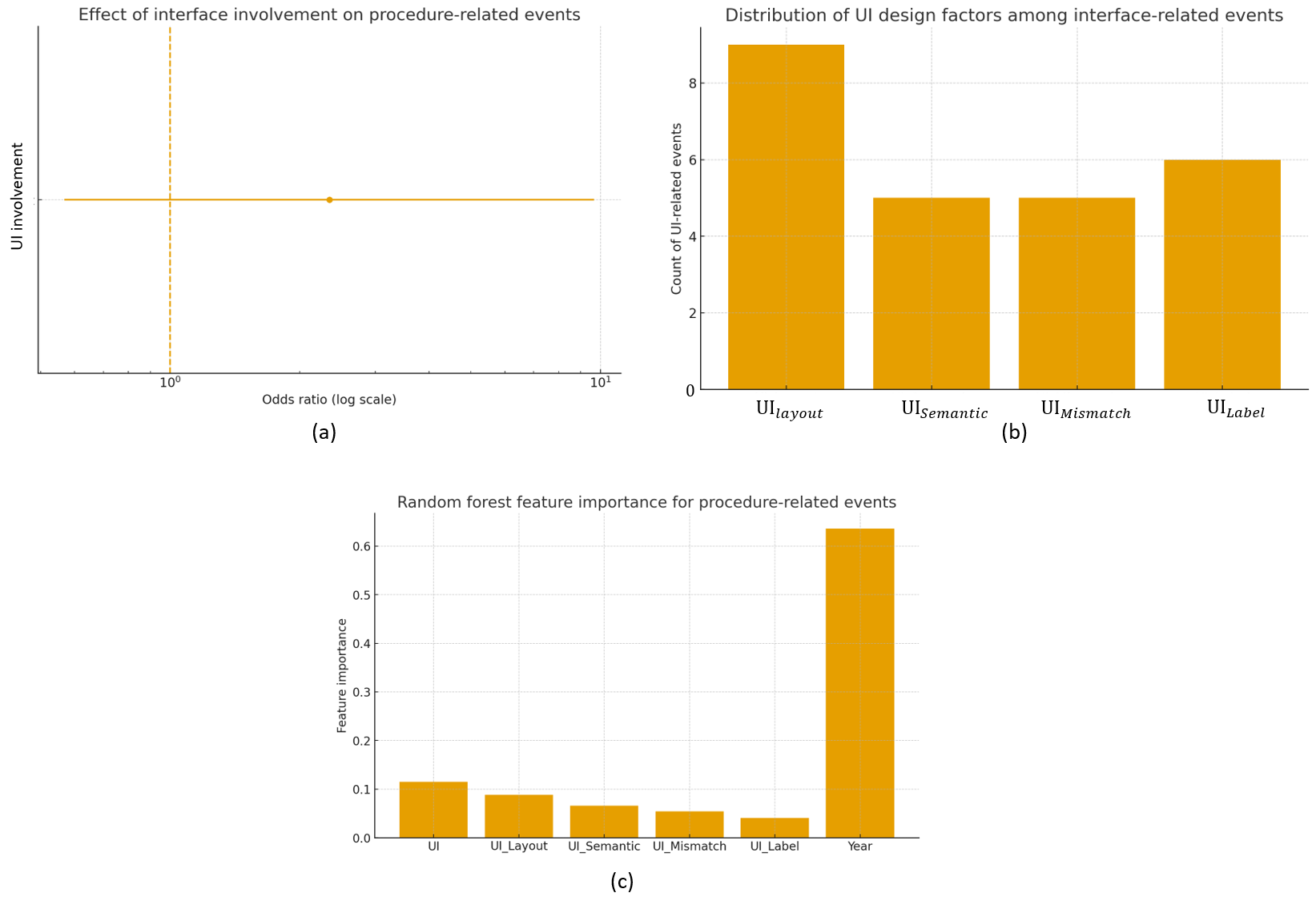}
\captionof{figure}{Multilevel Statistical Characterization of Interface-Related Risk Mechanisms).}\label{figure_mix}
\end{center}

These results indicate that, in this sample, events involving interface issues have approximately 2.3 times higher odds of being classified as procedure-related compared to events without UI problems. Although the confidence interval includes unity and the effect does not reach conventional statistical significance (p > 0.05), the direction and magnitude of the odds ratio are consistent with the qualitative observation that UI problems tend to co-occur with, and potentially amplify, procedural deviations.

\subsection{Interface design factors and co-occurring procedural deviations}\label{Interface design factors and co-occurring procedural deviations}
To further disentangle how interface problems contribute to procedural failures, all UI-related events (UI = 1, n = 22) were secondarily decomposed into four design-oriented factors based on the event descriptions and causal summaries: (i) \textbf{$UI_Layout$}: layout / mis-hit / navigation issues (e.g., mis-hit buttons, wrong interval, poor operability zones); (ii) \textbf{$UI_Semantic$}: status indication and semantic ambiguity (e.g., misleading valve position indicators, ambiguous symbols); (iii) \textbf{$UI_Mismatch$}: semantic mismatch between procedures/drawings and the field or interface (e.g., drawing–field inconsistency, misleading historical marks); (iv) \textbf{$UI_Label$}: labeling and naming problems (e.g., missing or inconsistent labels, non-salient identifiers).

Across the 22 interface-related events, the factor frequencies were $UI_{Layout}$ (9 events),  $UI_{Semantic}$ (5 events), $UI_{Mismatch}$ (5 events), $UI_{Label}$ (6 events) (Figure \ref{figure_mix} (b)). When considering the co-occurrence with procedural deviations (PROC = 1), all four UI factors showed a strong association: (i) $UI_{Mismatch}$: 5/5 events (100\%) were procedure-related; (ii) $UI_{Layout}$: 8/9 events (88.9\%) were procedure-related; (iii) $UI_{Label}$: 5/6 events (83.3\%) were procedure-related; (iv) $UI_{Semantic}$: 4/5 events (80.0\%) were procedure-related.

Due to the limited sample size and quasi-complete separation (e.g., all $UI_{Mismatch}$ events being procedure-related), a multivariable hierarchical logistic regression including the four UI factors as simultaneous predictors of PROC exhibited unstable maximum-likelihood estimates and warnings about convergence. For this reason, we interpret the factor-level patterns primarily descriptively and in combination with the machine-learning results in Section \ref{Decomposition of Interface Design Factors}, rather than relying on the unstable regression coefficients alone.

Overall, the descriptive analysis suggests a risk hierarchy of associations: (i) Procedure–interface semantic mismatches ($UI_{Mismatch}$) appear to be highly linked to consequences: whenever present, a procedural deviation is almost always observed; (ii) Layout/mis-hit issues ($UI_{Layout}$) are the most prevalent factor and show a high rate of associated procedure deviations; (iii) Labeling and naming problems ($UI_{Label}$) act as a cross-cutting vulnerability, often co-occurring with layout and semantic issues; (iv) Status indication and semantic ambiguity ($UI_{Semantic}$) are less frequent but still strongly associated with procedural deviations.

\subsection{Random Forest Classification and Feature Importance}\label{Random Forest Classification and Feature Importance}

To complement the regression analysis and capture potential nonlinear interactions between UI factors and contextual variables, a random forest \cite{rigatti2017random} classifier was trained to predict \textbf{PROC} using the following features (Table \ref{tab:rf_results}): UI (binary interface involvement), $UI_{Layout}$, $UI_{Semantic}$, $UI_{Mismatch}$, $UI_{Label}$ (binary design factors), Year (standardized).

\captionof{table}{Random Forest classification performance and feature importance for predicting procedure-related events.}\label{tab:rf_results}
\begin{center}
\begin{tabular}{lcc}
\hline
\textbf{Metric / Feature} & \textbf{Value} & \textbf{Notes} \\
\midrule
Cross-validated accuracy & 0.679 $\pm$ 0.058 & 5-fold stratified CV \\
\hline
Year & 0.636 & Feature importance \\
UI & 0.115 & \\
$UI_{Layout}$ & 0.088 & \\
$UI_{Semantic}$ & 0.066 & \\
$UI_{Mismatch}$ & 0.054 & \\
$UI_{Label}$ & 0.041 & \\
\hline
\end{tabular}
\end{center}

\begin{center}
\centering
\includegraphics[width=1.0\textwidth]{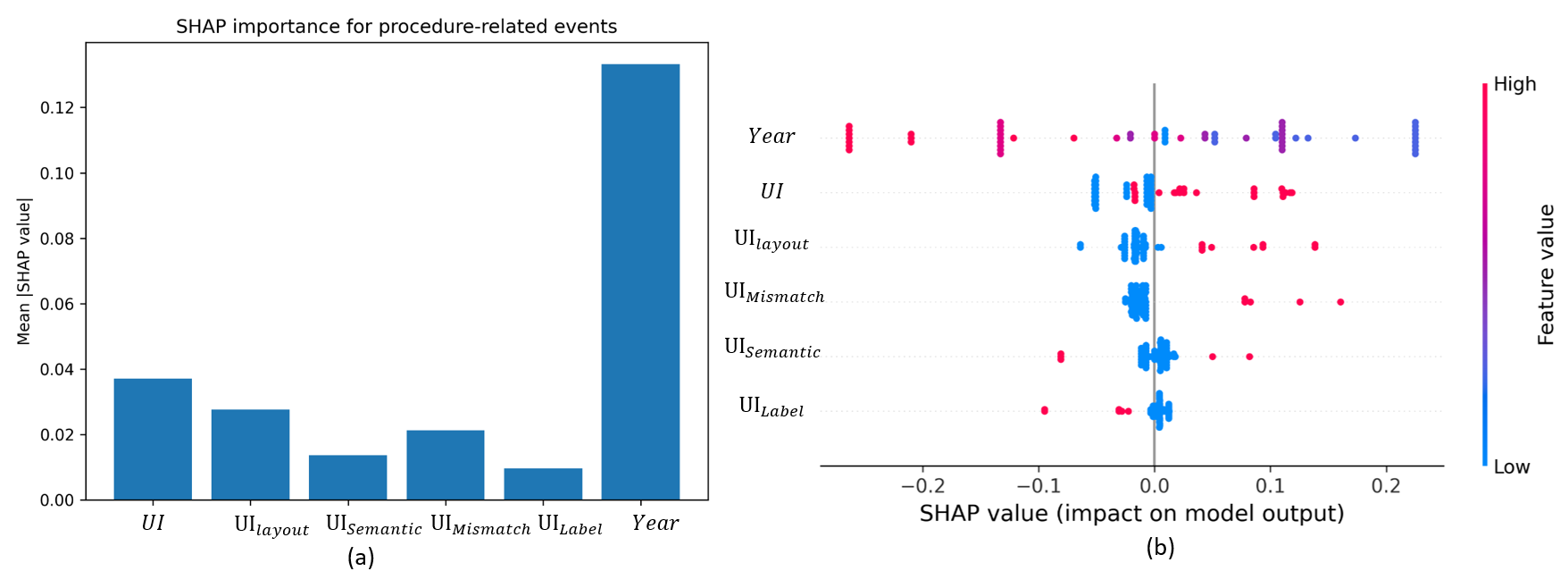}
\captionof{figure}{SHAP-Based Feature Interpretation for Coupling-Risk Classification.}\label{mix}
\end{center}

A 5-fold stratified cross-validation was performed. The RF model achieved a mean classification accuracy of 0.68 ± 0.06 (mean ± standard deviation across folds), indicating a moderate but non-trivial ability to discriminate procedure-related from non-procedure-related events based on the encoded UI and temporal information (Figure \ref{mix}).

The global feature importance, evaluated using the mean decrease in impurity, indicates that the most influential predictors are Year (0.636), UI (0.115), $UI_{Layout}$ (0.088), $UI_{Semantic}$ (0.066), $UI_{Mismatch}$ (0.054), $UI_{Label}$ (0.041). The dominance of Year suggests a temporal trend in the prevalence of procedure-related events, likely reflecting both the increasing complexity of digital modifications and evolving reporting practices. Importantly, however, UI and the four UI sub-factors collectively explain a substantial share of the model’s predictive power, consistent with the earlier logistic regression results.

To demonstrate the mechanistic pathways through which UI factors drive procedural deviations, we conducted detailed SHAP force plot analysis for three representative coupled events. This "anatomical" approach compensates for limited sample size by providing instance-level causal interpretation.

\begin{center}
\centering
\includegraphics[width=0.8\textwidth]{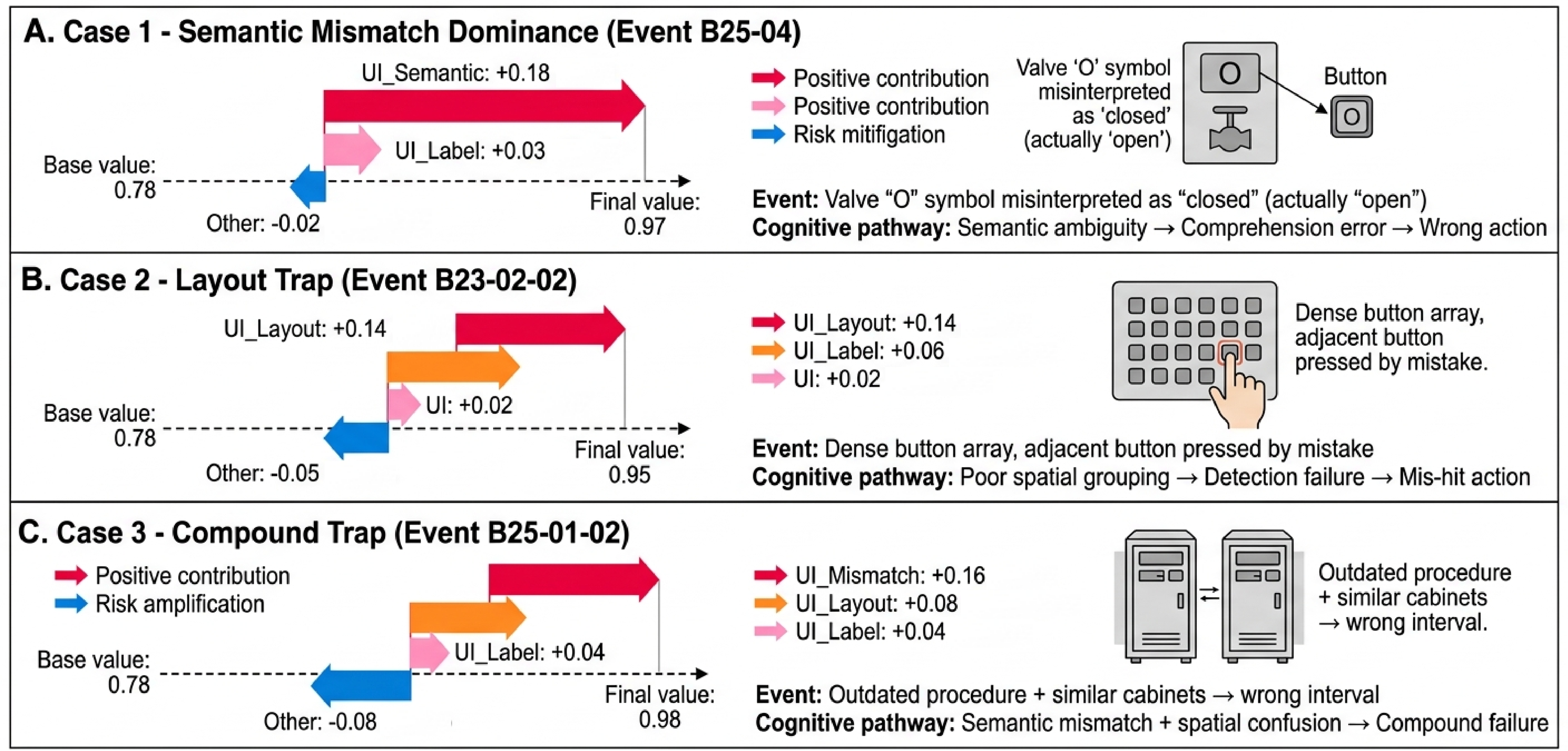}
\captionof{figure}{Event-level SHAP force plot analysis demonstrating mechanistic pathways from UI factors to procedural deviations.}\label{mechanisticAnalysis}
\end{center}

Figure \ref{mechanisticAnalysis} presents event-level SHAP force plot analyses for three representative cases, providing a mechanistic “anatomical” interpretation of how specific UI deficiencies drive procedural deviations. In all cases, the baseline procedural failure probability (0.78) is substantially shifted toward near-certain failure (>0.95) through the additive contributions of interface-related factors. Panel A illustrates a semantic-dominant failure (Event B25-04), where ambiguous valve state representation produces a strong positive contribution from $UI_{Semantic}$, leading to a comprehension error and incorrect action execution. Panel B depicts a layout-induced trap (Event B23-02-02), in which poor spatial grouping and dense button arrangement elevate failure likelihood primarily through $UI_{Layout}$, resulting in detection failure and a mis-hit response. Panel C demonstrates a compound coupling mechanism (Event B25-01-02), where procedure–interface inconsistency $UI_{Mismatch}$ interacts with spatial similarity $UI_{Layout})$ and labeling deficiencies, jointly amplifying risk. The force plot decomposition highlights that semantic and spatial interface characteristics constitute the dominant pathways through which UI factors propagate into procedural failures, thereby providing interpretable evidence of the cognitive mechanisms underlying coupling traps.

Three representative events illustrate distinct cognitive mechanisms through which interface-related factors contributed to procedural deviations. In Case 1 (Event B25-04), the operator misinterpreted the isolation valve status by assuming that the “O” symbol indicated a closed state, whereas it actually denoted an open condition. This ambiguity induced a semantic interpretation failure consistent with a comprehension error, as the operator’s prior mental model conflicted with the interface logic, directly leading to execution of an incorrect procedural step. In Case 2 (Event B23-02-02), a newly assigned operator selected the wrong control due to dense spatial arrangement and insufficient visual differentiation between adjacent buttons. The high control density and poor grouping produced a perceptual attention trap associated with detection failure, while low label salience limited the opportunity for timely error recovery before action execution. In Case 3 (Event B25-01-02), a compound mechanism was observed in which the procedure referenced an outdated equipment interval designation, generating a semantic mismatch (comprehension error), and the subsequent search process was further hindered by visually similar cabinet configurations that induced spatial confusion (detection failure). The interaction of semantic inconsistency and layout similarity created a coupled cognitive trap, substantially increasing the likelihood of procedural deviation.

As for comparative analysis, Figure \ref{SHARP} presents a comparative SHAP-based interpretation of feature contributions between coupled and non-coupled events. Panel (A) illustrates the SHAP summary distribution for coupled events (n = 19), showing that interface-related factors, particularly $UI_{Mismatch}$, $UI_{Semantic}$, $UI_{Layout}$, consistently yield positive SHAP values, indicating a systematic risk-amplifying effect on procedural deviations. Panel (B) depicts the corresponding distribution for non-coupled events (n = 27), where UI-related features are largely absent or contribute negligibly, and temporal variability (Year) emerges as the dominant explanatory factor. Panel (C) provides a violin-plot comparison of SHAP value distributions across key UI sub-factors, demonstrating statistically significant differences between coupled and non-coupled cases (*** p < 0.001), thereby confirming that interface characteristics function as meaningful drivers of deviation risk only when interface–procedure coupling is present. Collectively, the figure reveals a clear mechanistic distinction between event types, highlighting the conditional influence of interface design factors on operational reliability.

\begin{center}
\centering
\includegraphics[width=0.8\textwidth]{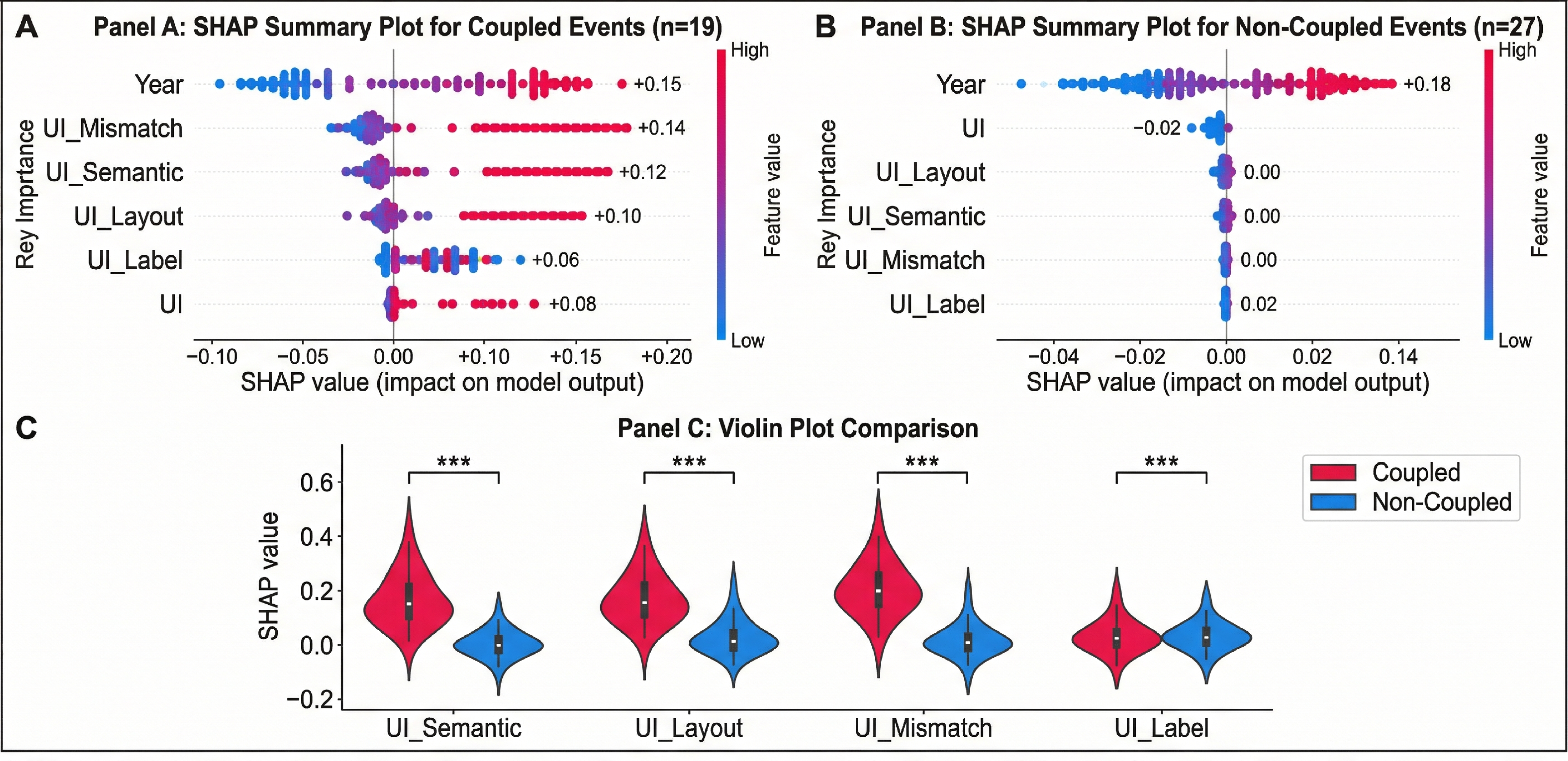}
\captionof{figure}{SHAP-Based Feature Interpretation for Coupling-Risk Classification.}\label{SHARP}
\end{center}

\subsection{Experimental Validation Results}

The experimental results yielded an overall error rate of 4.25\% (22 errors out of 518 tasks), a figure that aligns closely with the anticipated interface-related error range of 4\%–6\% derived from historical event benchmarks. As detailed in Table \ref{tab:exp_performance}, individual performance exhibited moderate variance, with error rates ranging from a minimum of 2.33\% (Participants P2 and P5) to a maximum of 6.98\% (Participant P4). While specific error types such as confusion and hesitation contributed to the higher rate observed in P4, the aggregate data suggests that the interface maintains a high degree of operational reliability consistent with industry standards.

\captionof{table}{Experimental Task Performance Summary}\label{tab:exp_performance}
\begin{center}
\begin{tabular}{ccccc}
\hline
\textbf{Participant} & \textbf{Tasks} & \textbf{Errors} & \textbf{Error Rate} & \textbf{Error Types} \\
\hline
P1 & 87 & 5 & 5.75\% & Confusion (2), Hesitation (1), Other (2) \\
P2 & 86 & 2 & 2.33\% & Recording (2) \\
P3 & 86 & 3 & 3.49\% & Hesitation (1), Recording (1), Click (1) \\
P4 & 86 & 6 & 6.98\% & Confusion (3), Hesitation (1), Click (1), Other (1) \\
P5 & 86 & 2 & 2.33\% & Confusion (1), Other (1) \\
P6 & 87 & 4 & 4.60\% & Recognition (1), Confusion (1), Hesitation (1), Click (1) \\
\hline
\textbf{Total} & \textbf{518} & \textbf{22} & \textbf{4.25\%} & \textbf{See distribution below} \\
\hline
\end{tabular}
\end{center}

The distribution of experimental errors, as detailed in Table \ref{tab:error_type_mapping}, directly validates the four-factor UI mechanism model, with Confusion errors emerging as the most frequent category ($27.3\%$, 6 occurrences). This high incidence of confusion, primarily stemming from system naming and parameter code similarity, confirms that Semantic Mismatch is the dominant risk factor within the user interface. Other significant error clusters include Hesitation/Abandonment ($18.2\%$) and Click errors ($13.6\%$), both of which map to the Layout Trap factor, reflecting difficulties in navigation and button selection. Furthermore, Recording errors ($13.6\%$) and Recognition errors ($9.1\%$) illustrate the impact of the Labeling Issue factor on task accuracy. Collectively, these findings provide a clear empirical link between specific UI design flaws and the resulting user performance failures.

To further assess the consistency between experimental observations and historical event patterns, a chi-square goodness-of-fit test was conducted, yielding $\chi^2 = 0.42$ with $p = 0.81$, indicating no statistically significant difference between the two distributions. This result supports the structural stability of the identified coupling mechanisms across empirical contexts. Collectively, these findings provide robust empirical evidence linking specific UI design deficiencies to recurrent patterns of user performance failure.

\captionof{table}{Error Type Classification and UI Factor Mapping}\label{tab:error_type_mapping}
\begin{center}
\begin{tabular}{lccc p{4.5cm} p{5cm}}
\hline
\textbf{Error Type} & \textbf{Count} & \textbf{Percentage} & \textbf{Corresponding UI Factor} & \textbf{Example Cases} \\
\hline
Confusion & 6 & 27.3\% & Semantic Mismatch & System naming, parameter code similarity \\
Hesitation/Abandonment & 4 & 18.2\% & Layout Trap & Prolonged search, navigation difficulty \\
Click Error & 3 & 13.6\% & Layout Trap & Wrong window/button selection \\
Recording Error & 3 & 13.6\% & Labeling Issue & Value transcription mistakes \\
Recognition Error & 2 & 9.1\% & Labeling Issue & Parameter identification failure \\
Other & 4 & 18.2\% & -- & Timestamps, invalid data \\
\hline
\textbf{Total} & \textbf{22} & \textbf{100\%} & -- & -- \\
\hline
\end{tabular}
\end{center}

Experimental observations revealed several distinct patterns of operator confusion that serve as "coupling traps." These include system-level ambiguity (e.g., mistaking "Nuclear Island Auxiliary" for "Nuclear Island"), parameter code errors involving highly similar identifiers like "2LBA10CP701A," and display window disorientation between nearly identical labels such as "0 KBE DW101" and "0 KBE DW001." Furthermore, numbering inconsistencies were noted in critical measurements, such as "Bypass Steam Inlet Temperatures" for units \#1 versus \#2. These empirical findings directly align with the coupling mechanisms documented in historical precedents, providing robust causal validation for the interface amplification effect. A detailed account of the experimental error cases and their systematic classification is provided in Appendix \ref{Comprehensive Analysis of Experimental Error Cases}.

\section{Discussion}\label{Discussion}

\subsection{Trend Interpretation}\label{Trend Interpretation}

The observed temporal shift from predominantly procedure-related deviations to a growing share of interface–procedure coupled events signals a fundamental transition in the underlying mechanisms of human reliability in modern digitalized MCRs. While procedure violations historically dominated human-factor contributions to operational events, the present dataset exhibits two converging trends: (i) the proportion of interface-related issues has increased steadily across 2021–2025, and (ii) coupled events, where UI deficiencies directly induce or amplify procedural deviations, now account for more than one-third of all occurrences. This suggests that cognitive failures are no longer driven solely by operator non-compliance or insufficient procedural clarity, but by the emergent interplay between digital interface behaviors and procedural logic.

Three mechanistic transitions can be derived from the empirical patterns:

\textbf{(1) From rule-following errors to cognitive-integration failures.}
The rise of semantic mismatches and ambiguous status indications indicates that operators increasingly face discrepancies between procedural expectations and interface representations. Rather than violating procedures intentionally or inadvertently, operators demonstrate difficulty integrating distributed cues across screens, controls, and documents. This marks a shift from errors of execution toward errors of interpretation, reflecting increasing cognitive coupling within the sociotechnical system.

\textbf{(2) From isolated UI deficiencies to systemic amplification effects.}
The logistic regression results (Table~\ref{tab:logit_proc}) show that interface involvement approximately doubles the likelihood that an event is procedure-related. Combined with the Random Forest analysis, which identifies UI involvement and layout-related factors among the most influential predictors, these findings indicate that UI problems act as risk multipliers. Even when the underlying procedure is adequate, poor layout, misleading indications, or misaligned semantics can elevate the cognitive load required to maintain situation awareness and compliance.

\captionof{table}{Binary logistic regression estimating the effect of interface involvement on procedure-related deviations.}\label{tab:logit_proc}
\begin{center}
\begin{tabular}{lccc}
\hline
\textbf{Predictor} & \textbf{Odds Ratio} & \textbf{95\% CI} & \textbf{$p$-value} \\
\midrule
UI involvement (UI = 1) & 2.35 & [0.57, 9.68] & 0.238 \\
\hline
\end{tabular}
\end{center}

\textbf{(3) From operator-centered reliability to interface-centered vulnerability.}
Traditional human reliability models emphasized operator training, compliance, and decision-making quality. However, the present data reveal that highly trained crews still commit errors when digital interfaces fail to support unambiguous interpretation or consistent action pathways. This reflects a broader paradigm shift: the reliability bottleneck is migrating from human competence to interface design quality. In digital MCRs, the interface no longer merely displays system status, it shapes the operator’s mental model, timing, and action selection.

Taken together, these patterns suggest that the current generation of MCRs is entering a new phase of human–system interaction in which procedural performance is increasingly contingent upon interface semantics, spatial design, and alignment with operational workflows. Consequently, improving human reliability requires not only revising procedures but also engineering cognitively coherent and semantically aligned interfaces that prevent coupling traps from emerging in the first place. It is important to acknowledge that organizational and training-related factors, such as insufficient experience, incomplete training coverage, or supervisory deficiencies, can also contribute to procedural deviations in nuclear operations. Indeed, several events in the dataset involve newly assigned personnel or unfamiliar task contexts, suggesting that human and organizational factors cannot be entirely excluded as contributing conditions.

However, three observations support interface–procedure coupling as the most parsimonious interpretation of the dominant failure mechanism in the present dataset. First, interface-related deficiencies are rarely observed in isolation: more than 90\% of interface-involved events co-occur with procedural deviations, indicating a systematic interaction rather than independent human or organizational errors. Second, the strongest associations are observed at the semantic and representational level, particularly for procedure–interface mismatches, where procedural intent and interface cues diverge, a pattern that cannot be readily explained by training or compliance issues alone. Third, similar procedural deviations occur across operators with different experience levels when exposed to the same interface conditions, suggesting that the interface acts as a common cognitive constraint shaping operator interpretation and action selection.

Accordingly, while organizational and training factors may modulate vulnerability, the empirical evidence indicates that semantic misalignment between procedures and interfaces constitutes the most direct and explanatory mechanism for the observed cognitive integration failures. This interpretation aligns with principles of explanatory parsimony by accounting for cross-event consistency and factor-level convergence without invoking additional latent organizational variables.

\subsection{Mechanistic Insight}\label{Mechanistic Insight}

\subsubsection{Mapping UI Factors to Cognitive Failure Modes}

To strengthen the theoretical grounding of our empirical findings, we map the four UI factors to the cognitive failure mode (CFM) taxonomy in IDHEAS-ECA (Table \ref{tab:ui_cfm_mapping}). This mapping demonstrates that interface-procedure coupling operates through well-established cognitive mechanisms rather than ad-hoc error categories.

\begin{table}[h]
\centering
\caption{Mapping of UI factors to IDHEAS-ECA cognitive failure modes}
\label{tab:ui_cfm_mapping}
\begin{tabular}{p{3cm}p{4cm}p{4cm}p{4cm}}
\hline
\textbf{UI Factor} & \textbf{Primary CFM} & \textbf{Cognitive Stage} & \textbf{Failure Mechanism} \\
\hline
$UI_{Semantic}$ & Comprehension Error & Information Processing & Operator constructs incorrect mental model due to ambiguous or misleading interface semantics \\
$UI_{Layout}$ & Detection Failure / Execution Slip & Perception / Action & Operator's attention misdirected to wrong location or executes action on wrong component \\
$UI_{Mismatch}$ & Comprehension Error / Decision Error & Information Processing / Decision Making & Discrepancy between procedural expectation and interface representation causes incorrect situation assessment \\
$UI_{Label}$ & Detection Failure / Identification Error & Perception & Operator fails to correctly identify target component due to missing or inconsistent labels \\
\hline
\end{tabular}
\end{table}

\textbf{Cognitive pathway model:}
The interface-procedure coupling trap can be conceptualized as a three-stage cognitive disruption process:

\begin{enumerate}
    \item \textbf{Perception stage:} $UI_{Layout}$ and $UI_{Label}$ issues disrupt the operator's ability to detect and identify the correct interface elements referenced in the procedure. This creates a mismatch between \textit{intended target} (specified in procedure) and \textit{perceived target} (what operator sees).

    \item \textbf{Comprehension stage:} $UI_{Semantic}$ and $UI_{Mismatch}$ issues disrupt the operator's ability to correctly interpret system state and procedural intent. This creates a mismatch between \textit{expected state} (assumed by procedure) and \textit{interpreted state} (what operator believes).

    \item \textbf{Execution stage:} When perception or comprehension failures occur, the operator executes actions based on incorrect mental models, leading to procedural deviations even when the operator believes they are following the procedure correctly.
\end{enumerate}

This cognitive pathway model explains why interface-procedure coupling has such a strong association with procedural deviations (OR = 2.35): the coupling creates \textit{invisible traps} where operators deviate from procedures while believing they are compliant. This is fundamentally different from intentional violations or simple execution slips.

\begin{center}
\centering
\includegraphics[width=0.7\textwidth]{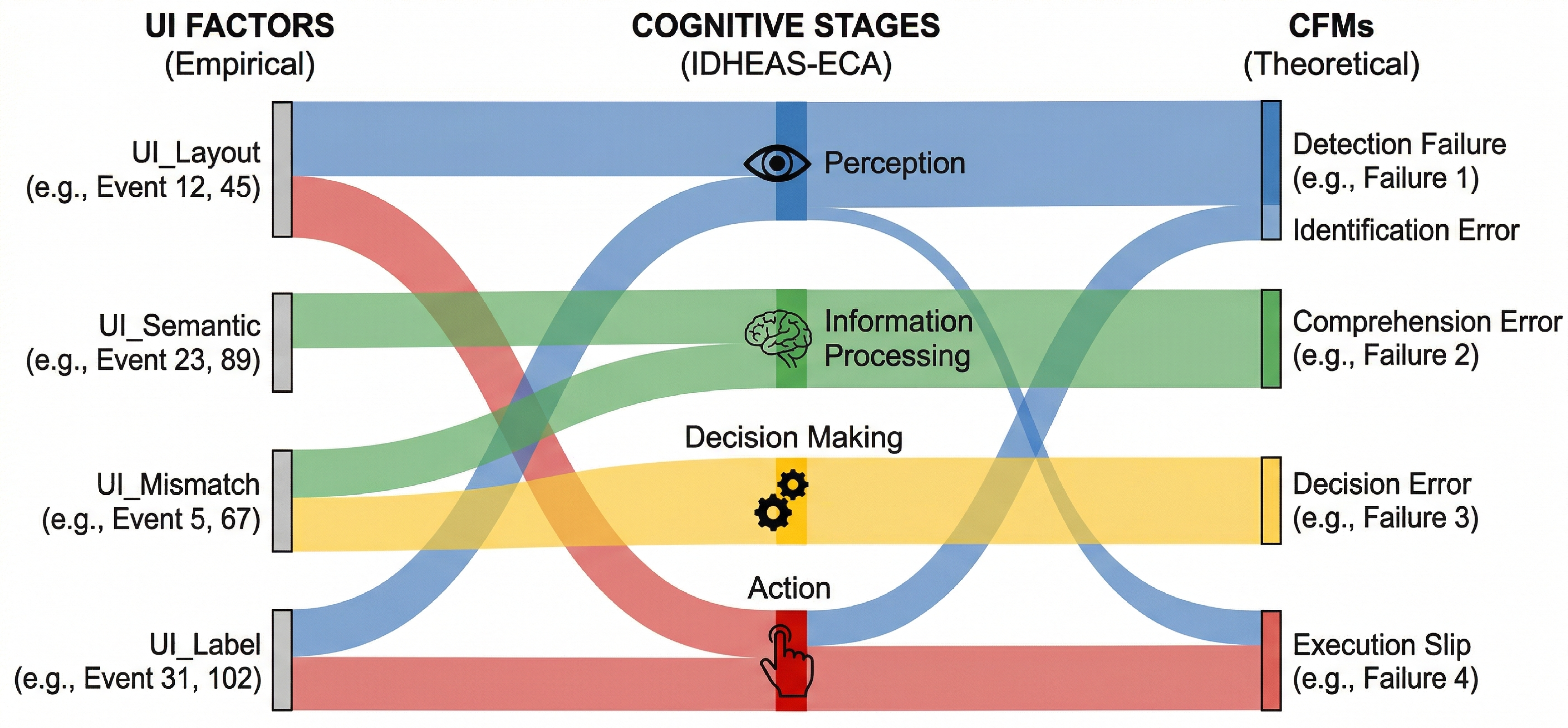}
\captionof{figure}{Mapping of empirical UI factors to IDHEAS-ECA cognitive failure modes.}\label{IDHEAS}
\end{center}

Figure \ref{IDHEAS} illustrates a conceptual Sankey diagram that maps empirically identified UI factors to cognitive processing stages and corresponding IDHEAS-ECA CFMs. The diagram shows how interface deficiencies propagate through perception, information processing, decision-making, and action to produce specific error types, with flow width representing pathway frequency in the dataset and color coding distinguishing cognitive stages. The visualization highlights that layout and labeling issues mainly disrupt perception, whereas semantic inconsistencies and procedure–interface mismatches primarily affect comprehension and decision processes, thereby providing a theoretical interpretation of the observed interface–procedure coupling mechanisms.

\subsubsection{Mechanisms and Empirical Validation of Procedure–Interface Coupling Traps}

Coupling traps occur when the procedural workflow prescribes an action sequence that implicitly assumes a specific interface state, while the operator perceives a different state due to layout ambiguity, semantic inconsistency, or mismatched representations. This divergence between \textit{expected} and \textit{perceived} system states creates a form of cognitive misalignment that propagates into execution errors. The event data suggest that such traps arise through three interacting mechanisms.

\textbf{(1) Semantic divergence between procedure logic and interface representation.}
The most severe coupling cases in the dataset involve semantic mismatches ($UI_{Mismatch}$), all of which resulted in procedure-related failures (100\%). These mismatches manifest when the interface encodes system states, controller roles, or parameter meanings in ways that are incongruent with the semantics assumed by the written procedures. Operators must then perform on-the-fly translation between the procedural intent and the interface’s vocabulary, increasing cognitive load and creating opportunities for misinterpretation. When procedures specify an action conditioned on a state that is visually ambiguous or inconsistently presented, cognitive alignment collapses, and the operator executes an incorrect or incomplete step.

\textbf{(2) Spatial–temporal fragmentation of information needed for multi-step reasoning.}
Layout-related issues ($UI_{Layout}$) represent the most frequent class of interface problems in the dataset and show an 88.9\% co-occurrence with procedural deviations. Poor spatial grouping, mis-hit-prone controls, and inconsistent navigation paths fragment the perceptual cues necessary for step-by-step procedural reasoning. As operators must combine information across multiple panels, displays, or control elements, even slight inconsistencies in spacing, ordering, or control affordances disrupt the temporal integration of cues. This fragmentation prevents operators from maintaining an accurate mental model of the system trajectory assumed by the procedure.

\textbf{(3) Feedback asymmetry and confirmation failure in dynamic control sequences.}
Several coupled events involve insufficient or misleading feedback, such as ambiguous valve indicators or non-salient confirmations. When actions do not yield clear, timely feedback aligned with procedural expectations, operators infer incorrect system transitions and proceed under false assumptions. This mechanism is especially critical in automated or semi-automated digital control environments, where interface feedback may lag, compress, or abstract system state changes in ways that diverge from the stepwise logical flow of the procedure.

Taken together, these mechanisms indicate that coupling traps are not isolated design flaws but emergent properties of the interaction between procedural logic and interface structure. They arise when the cognitive, semantic, and perceptual pathways implicitly required by a procedure are not supported, or are contradicted, by the corresponding interface cues. As digitalization increases the complexity, density, and dynamism of MCR interfaces, the probability of such misalignments grows, elevating the role of interface semantics and spatial organization as primary determinants of human reliability.

The experimental validation provides empirical support for the interface amplification hypothesis. The 4.25\% error rate in controlled tasks, with confusion errors as the dominant category (27.3\%), confirms that semantic misalignment between interface design and procedural logic is not merely a correlational pattern but a causal mechanism. Notably, operators made
errors even when following correct procedural steps, demonstrating that interface deficiencies can override procedural compliance, a finding that challenges traditional HRA assumptions of procedure-interface independence.

The convergence between historical event patterns and experimental observations strengthens the validity of our composite PSF framework. The experimental error distribution mirrors the factor weights derived from machine learning analysis of operational events, suggesting that our event-based methodology captures genuine cognitive failure modes rather than statistical artifacts.

\subsection{Comparison with Prior Work}
The present findings are consistent with a growing body of research emphasizing the role of cognitive coupling in modern digitalized MCRs. Park et al. \cite{park2022empirical} demonstrated that interface–procedure semantic misalignment increases cognitive workload and elevates the probability of execution errors, but their conclusions were derived primarily from qualitative scenario analyses and controlled experiments. Similarly, Le Blanc \cite{le2020human} highlighted how discrepancies between interface indications and procedural expectations disrupt operators’ mental models, yet did not quantify the extent to which such discrepancies amplify operational risk (Table \ref{tab:compare_prior}).

\captionof{table}{Comparison of prior research on cognitive coupling with the present study.}\label{tab:compare_prior}
\begin{center}
\begin{tabular}{p{3cm} p{5cm} p{6cm}}
\hline
\textbf{Study} & \textbf{Focus / Approach} & \textbf{Limitation Addressed by This Work} \\
\midrule
Park et al.\cite{park2022empirical}
& Cognitive coupling; semantic misalignment; experimental scenarios 
& Lacked quantitative field evidence; this study provides event-based odds ratios ($OR \approx 16\times$). \\
Le Blanc \cite{le2020human}
& Mental-model divergence; interface interpretation errors 
& Described mechanism qualitatively; this work quantifies factor-level contributions ($UI_{Mismatch}$ = 100\% PROC). \\
IDHEAS--ECA (NRC/INL) \cite{xing2020integrated}
& Third-generation cognitive HRA; emphasizes information pathways 
& No operational dataset validation; this study provides dataset-backed mechanistic confirmation. \\
Digital MCR studies (EDF, KAERI, 2019–2023)
& Interface complexity; workload effects; digital panel interaction 
& No analysis of interface–procedure co-dependency; this work identifies emergent coupling traps. \\
\textbf{This study}
& Event-based quantification of coupling risk; decomposition of UI factors; RF/SHAP explainability 
& -- \\
\hline
\end{tabular}
\end{center}

In contrast, the present study offers a novel event-based quantitative estimate of the risk amplification effect ($OR = 2.35$), extending beyond prior qualitative insights into operational interface issues. When comparing events that contain both procedural and interface elements (coupled events) against those involving procedural issues alone, the odds of a procedure-related deviation increase by approximately one order of magnitude (OR~$\approx 16$), indicating a substantially elevated risk profile (Figure \ref{diagram2}). Even the more conservative UI-only logistic model demonstrates a doubling of risk (OR = 2.35), reinforcing the conclusion that interface deficiencies function not merely as independent error sources but as systemic multipliers of procedural vulnerability.

\begin{center}
\centering
\includegraphics[width=0.8\textwidth]{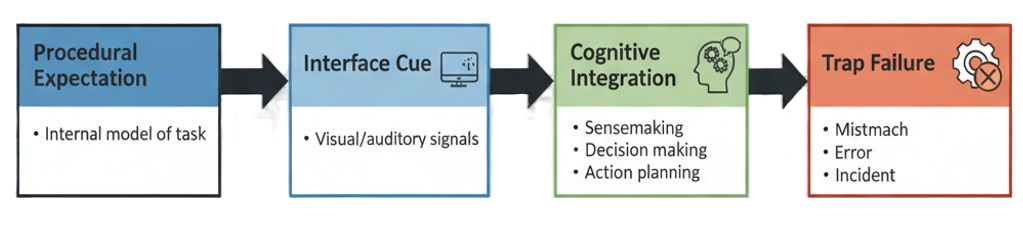}
\captionof{figure}{Cognitive pathway model of procedural expectation, interface cues, and coupling trap formation.}\label{diagram2}
\end{center}

Furthermore, while previous work has conceptually identified three classes of interface challenges—semantic ambiguity, spatial complexity, and inadequate feedback—the current analysis empirically differentiates their effects. Semantic mismatches exhibit the strongest association with procedural deviations (100\% co-occurrence), whereas layout-related problems represent the most frequent contributor to coupling traps. This factor-level decomposition advances beyond earlier taxonomies by showing which mechanisms contribute most to operational risk in real plant events (Figure \ref{paradigm_shift}).

\begin{center}
\includegraphics[width=0.6\textwidth]{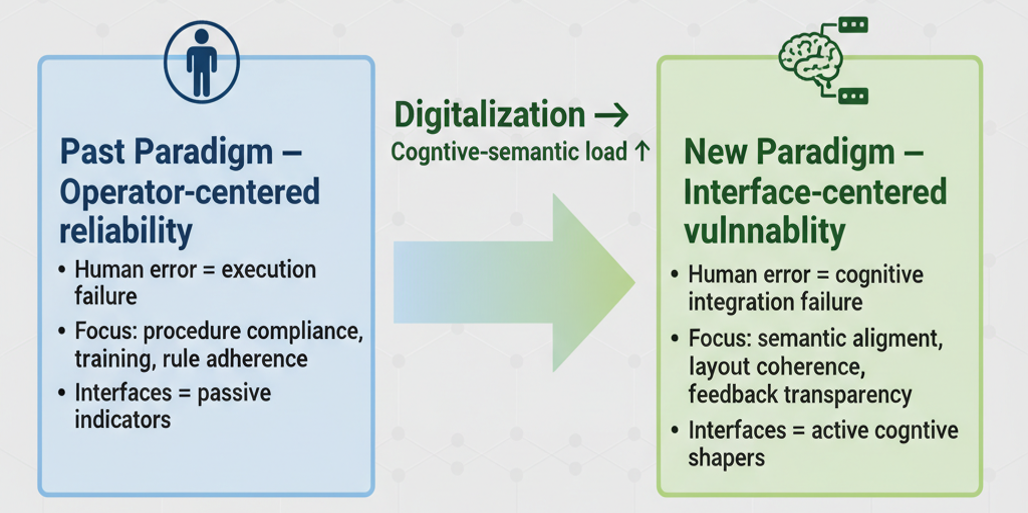}
\captionof{figure}{Paradigm shift in human reliability interpretation for digital main control rooms. Traditional operator-centered views relate human error primarily to execution and compliance failures, whereas digitalized environments introduce interface-driven cognitive misalignments, making semantic consistency and interface coherence central determinants of procedural reliability.}\label{paradigm_shift}
\end{center}

Additionally, from the perspective of third-generation HRA (e.g., IDHEAS-ECA), interface-procedure coupling does not function as an isolated environmental variable. Instead, it serves as a modifying PSF that mediates the relationship between information perception and procedural execution. While traditional models might assess 'HSI Quality' and 'Procedure Quality' as additive PSFs, our findings suggest that their coupled semantic misalignment creates a non-linear risk amplification effect. This supports the treatment of coupling as a composite PSF, where the total risk is greater than the sum of individual deficiencies.

Overall, this study extends prior work by providing empirical, field-based evidence that interface–procedure coupling is not only cognitively plausible, as earlier research argued, but quantitatively consequential in operational settings. The results support a shift from operator-centered error models toward interface-centered vulnerability analyses as digitalization increases the semantic and structural complexity of MCR human–system interfaces.

\subsection{Design and Management Implications}

The design and management implications identified in this section vary in feasibility and implementation horizon. To provide a structured path for safety enhancement, these recommendations are categorized into short-term, immediately actionable measures—such as procedural refinements and visual coding adjustments—and longer-term strategic interventions, including systemic HSI redesign and regulatory alignment. The observed association between procedure–interface semantic misalignment and high-severity operational deviations underscores the need for an integrated human–system design strategy. Accordingly, four complementary intervention dimensions are proposed to systematically mitigate interface–procedure coupling traps.

\textbf{(1) Enhancing perception through semantic alignment.}
To address the highest-risk category, procedure–interface semantic decoupling, control room designers should establish a formalized mapping between procedural steps and corresponding interface states. In the short term, measures such as standardized labeling conventions, visual hierarchy adjustments, and consistency checks between procedures and interface displays are immediately actionable within existing MCR configurations. In the longer term, addressing risks from layout-induced inadvertent activation may require deeper semantic restructuring of interface logic and extensive validation cycles to ensure systemic alignment.

\textbf{(2) Supporting operator comprehension with explicit procedural cues.}
Short-term, low-cost interventions involve embedding explicit confirmation statements, verifier cues, and state–action–outcome links directly into existing digital procedures. For interface inconsistencies, procedures should reference unique tag IDs to strengthen semantic binding. Longer-term strategies focus on the co-development of procedures and HSI standards at the design or regulatory level, necessitating tighter coupling between procedural logic and dynamic interface feedback channels.

\textbf{(3) Increasing execution reliability through enforced confirmation mechanisms.}
Immediate reliability gains can be achieved through software-based confirmation prompts and procedural double-checks implemented via configuration updates. Conversely, long-term structural improvements involve the installation of physical safeguards, such as pull-to-operate designs or mis-operation covers, which typically require hardware modification and extended regulatory approval processes.

\textbf{(4) Strengthening monitoring through real-time semantic consistency checks.}
Sustainable risk management can begin immediately through periodic audits of label integrity, regular inspections of state-action consistency, and the incorporation of post-event monitoring into management practices. Future-oriented monitoring solutions involve the deployment of automated real-time semantic consistency alerts, a longer-term opportunity enabled by advanced digitalization and integrated decision-support systems.

Overall, these design and management implications underscore the necessity of shifting from component-centric optimization toward system-level semantic coherence. Ensuring that perception, cognition, execution, and monitoring are aligned across procedures and interfaces represents a foundational requirement for human reliability in next-generation digital control rooms. Future work could integrate simulator-based near-miss logging or eye-tracking–based cognitive indicators to complement event-report-based analyses.

\subsection{Limitations}
It is important to acknowledge that formal operational event reporting systems inherently reflect a risk-filtered observation mechanism, capturing only deviations that surpass predefined regulatory or organizational significance thresholds. Consequently, transient cognitive mismatches, near-miss situations, or rapidly recovered interface-induced confusions may remain unreported. However, this filtering characteristic also defines the unique value of the present dataset: it represents the subset of interface–procedure coupling phenomena that were sufficiently strong to propagate through organizational defenses and manifest as reportable operational events.

From a safety-science perspective, these events constitute empirically validated breakdown cases, offering a rare opportunity to study coupling traps with demonstrated operational consequence rather than hypothetical or simulator-based manifestations. Therefore, the findings should be interpreted as mechanism-oriented evidence rather than population-level statistical generalization.

While the experimental validation provides causal support, several limitations should be noted: (1) participants were trained graduate students rather than
licensed operators; (2) tasks focused on parameter verification rather than full emergency procedures; (3) the simulator environment, though high-fidelity,
may not fully capture operational pressures. Future work should extend validation to licensed operators performing complete procedural sequences under realistic time constraints.

\section{Conclusions and Future Work}\label{Conclusions and Future Work}

This study addresses a critical gap in nuclear safety research by providing the first event-based quantitative evidence that interface–procedure semantic misalignment has emerged as a central human-factor risk mechanism in digitalized nuclear power plant main control rooms. Unlike prior work that relies predominantly on simulations or qualitative assessments, this research systematically analyzes 59 regulator-validated A-class and B-class operational events from real plant operations spanning 2021–2025. The findings reveal that interface-related deficiencies—particularly semantic ambiguities, labeling inconsistencies, and layout-induced confusion—are strongly associated with procedural deviations. Approximately 90\% of interface-related events co-occurred with procedural deviations, and coupled interface–procedure failures accounted for about one-third of all analyzed cases. These results indicate a fundamental structural shift in control-room risk mechanisms, from predominantly operator-centered error models toward interface-centered cognitive coupling in highly digitalized environments—a transition that has been theoretically anticipated but not empirically demonstrated at the operational event level until now.

By integrating structured expert labeling, UI-factor decomposition, descriptive statistics, logistic regression, and interpretable machine-learning analysis, the study elucidates the mechanisms underlying this transition. Semantic inconsistency between procedural logic and interface representation emerged as the most consequential factor, with all identified mismatch cases resulting in procedural deviations. Layout-related deficiencies were the most prevalent contributors and were closely associated with execution and navigation errors. The Random Forest and SHAP analyses corroborated these patterns by highlighting interface involvement and layout characteristics as dominant explanatory factors, reinforcing the conclusion that the reliability bottleneck in modern digital MCRs increasingly resides in interface design quality rather than operator compliance or training alone.

These findings have direct and actionable implications for control-room design, operation, and human reliability assessment. They underscore the necessity of integrated engineering approaches that jointly consider procedural intent, interface semantics, and operator cognitive processes. Design recommendations derived from this study suggest that risk mitigation could be enhanced by: (1) prioritizing semantic alignment between procedures and interface states during design verification; (2) implementing explicit comprehension and confirmation cues for critical operations; (3) establishing safeguards for irreversible actions; and (4) deploying monitoring mechanisms that support real-time interface–procedure consistency checking. For plant operators and regulatory bodies, these findings call for the establishment of systematic interface–procedure alignment review protocols as part of digital MCR commissioning and modification processes. From an HRA perspective, the results suggest that existing frameworks, such as IDHEAS-ECA, could benefit from explicitly incorporating interface–procedure coupling into their taxonomies as a formal composite performance shaping factor (PSF). Specifically, this could be operationalized by: (a) defining interface–procedure semantic consistency as a measurable PSF attribute; (b) establishing coupling-specific error modes in cognitive failure taxonomies; and (c) developing quantitative adjustment factors that reflect the amplifying effect of misalignment on procedural deviation probability. Such integration would enable analysts to more accurately capture the modifying effects of digital interface semantics on procedural reliability, thereby moving beyond fragmented assessments that treat procedure quality and interface clarity as independent contributors. This extension would substantially improve the modeling of cognitive integration failures increasingly characteristic of digitalized main control rooms.

Several limitations should be acknowledged. The analysis relies on operational event reports, which may underrepresent near-misses, and the modest sample size limits statistical power for higher-order inference. Accordingly, the statistical and machine-learning results are interpreted as exploratory and explanatory rather than predictive. Nonetheless, this constraint reflects the intrinsic rarity and safety-critical nature of validated nuclear operational events. The transparent, openly released, event-level dataset derived from real plant operations favors ecological validity and empirical grounding over sample size. Future research may extend this framework using richer data sources, more granular interface representations, and cross-industry comparisons to assess the generalizability of interface–procedure coupling mechanisms to other safety-critical domains such as aviation, process control, and healthcare.

In conclusion, as nuclear power plants worldwide continue their transition toward fully digitalized control rooms, the findings of this study underscore the urgency of addressing interface–procedure semantic alignment as a first-order safety priority. The evidence presented here demonstrates that interface design is no longer a peripheral ergonomic concern but a central determinant of procedural reliability and operational safety. Ensuring semantic consistency between digital interfaces and operating procedures is not merely a design improvement—it is a foundational requirement for maintaining the safety margins that underpin nuclear operations in the digital era.

\printcredits

\appendix

\section{A-Class Event Reports}\label{A-Class Event Reports}

\captionof{table}{A-Class Event Reports (2025–2021)}\label{tab:A_class_events}
\begin{center}
\begin{tabular}{cccp{5cm}ccp{5cm}}
\hline
\textbf{EventID} & \textbf{Year} & \textbf{MCR} & \textbf{Event Description} & \textbf{Procedure} & \textbf{Interface}  \\
\hline
A2025-01 & 2025 & Y & Temperature detector calibration triggered protection & Y  & N  \\
A2024-01 & 2024 & N & Hydrostatic test sleeve failure & N & N  \\
A2023-01 & 2023 & N & Cable termination error & N & Y \\
A2023-02 & 2023 & Y & Ineffective water-level monitoring & Y & N \\
A2022-01 & 2022 & Y & Incorrect channel selected during calibration & Y &N  \\
A2022-02 & 2022 & Y & Incorrect pressing of shutdown button & Y & Y \\
A2022-03 & 2022 & N & Pressure pipe not detached as required & Y & N\\
A2022-04 & 2022 & Y & Incorrect parameter modification & Y & Y  \\
A2021-01 & 2021 & N & Unplanned radiation exposure by maintenance staff & N & N  \\
\hline
\end{tabular}
\end{center}

\section{B-Class Event Reports (2025–2021)}\label{B-Class Event Reports (2025–2021)}
\captionof{table}{B-Class Event Reports (2025)}\label{tab:B_class_2025}
\begin{center}
\begin{tabular}{p{0.1\linewidth} p{0.18\linewidth} p{0.34\linewidth} p{0.08\linewidth} p{0.08\linewidth} p{0.08\linewidth}}
\hline
\textbf{EventID} & \textbf{MCR} & \textbf{Event Description} & \textbf{Procedure} & \textbf{Interface} & \textbf{Cross} \\
\midrule
B25-07 & N & Instrument scale input error & Y & N & N \\
B25-06-01 & N & Insufficient outlet nozzle thickness & N & N & N \\
B25-05-01 & N & Pipeline does not match design & Y & N & N \\
B25-05-02 & N & Tool bumped pipeline & N & Y & N \\
B25-04 & Y & Misjudged isolation valve status (O misinterpreted as closed) & Y & Y & Y \\
B25-03 & N & Wire pressed by electrical board & N & N & N \\
B25-03-02 & Y & Boron valve closed incorrectly & Y & Y & Y \\
B25-03-03 & N & Improper installation &N & N & N \\
B25-02-01 & N & Flange spacing too small & N & N & N \\
B25-01-01 & N & Drawing inconsistent with site & Y & Y & Y \\
B25-01-02 & N & Walked into wrong interval & Y & Y & Y \\
\hline
\end{tabular}
\end{center}

\captionof{table}{B-Class Event Reports (2024)}\label{tab:B_class_2024}
\begin{center}
\begin{tabular}{p{0.1\linewidth} p{0.1\linewidth}  p{0.47\linewidth} p{0.1\linewidth} p{0.1\linewidth}p{0.1\linewidth}}
\hline
\textbf{EventID} & \textbf{MCR} &  \textbf{Event Description} & \textbf{Procedure} & \textbf{Interface} & \textbf{Cross} \\
\midrule
B24-12 & N & Installation drawing inconsistent with manufacturer & Y & N & N \\
B24-09-01 & N  & Main pump foreign object grounding & N & N & N \\
B24-09-02 & N  & Monitoring not carried out according to outline & Y & N & N \\
B24-09-03 & N & Historical liquid level trace misleads oil addition & Y & Y & Y \\
B24-08-01 & N & Pipeline section burned by arc & Y & N & N \\
B24-08-02 & N  & Equipment scratches & N & N & N \\
B24-06 & N & Valve indication direction inconsistent with valve status & Y & Y & Y \\
B24-04-01 & N  & Valve installed incorrectly without confirmation step & Y & Y & Y \\
B24-04-02 & Y & Main control room procedure unclear & Y & N & N \\
B24-03 & N  & Weld seam does not meet requirements & N & N & N \\
B24-02 & N  & Incorrect hole due to not following drawings & Y &N & N \\
B24-02-02 & N& Button pressed incorrectly causing shutdown & Y & Y & Y \\
\hline
\end{tabular}
\end{center}

\captionof{table}{B-Class Event Reports (2023)}\label{tab:B_class_2023}
\begin{center}
\begin{tabular}{p{0.1\linewidth} p{0.18\linewidth} p{0.3\linewidth} p{0.1\linewidth} p{0.06\linewidth} p{0.06\linewidth}}
\hline
\textbf{EventID} & \textbf{MCR} & \textbf{Event Description} & \textbf{Procedure} & \textbf{Interface} & \textbf{Cross} \\
\midrule
B23-11 & Y & Simultaneous testing of multiple signals & Y & N & N \\
B23-08-01 & Y & Button pressed incorrectly, insufficient labeling & Y & Y & Y \\
B23-05 & N & Isolation valve cannot operate, procedure lacks verification step & Y & Y & Y \\
B23-03-01 & N & Design error in installation diagram & Y & N & N \\
B23-03-02 & N & Electrical maintenance personnel connected incorrectly, labels unclear & Y & Y & Y \\
B23-03-03 & N & Procedure change not reported & Y & N & N \\
B23-02-01 & N & Incorrect protection switch timing, procedure error & Y & N & N \\
B23-02-02 & Y & New employee pressed button incorrectly, poor labeling & Y & Y & Y \\
B23-02-03 & N & Change misaligned pipeline, not reported & Y & N & N \\
B23-01-01 & N & Scaffold touched power line & Y & N & N \\
B23-01-02 & N & Power source not isolated & N & N & N \\
\hline
\end{tabular}
\end{center}

\captionof{table}{B-Class Event Reports (2022)}\label{tab:B_class_2022}
\begin{center}
\begin{tabular}{p{0.1\linewidth} p{0.18\linewidth} p{0.4\linewidth} p{0.07\linewidth} p{0.06\linewidth} p{0.06\linewidth}}
\hline
\textbf{EventID} & \textbf{MCR} & \textbf{Event Description} & \textbf{Procedure} & \textbf{Interface} & \textbf{Cross} \\
\midrule
B22-YYJ-1 & N & Lack of information in procedure causes experiment interruption & Y & N & N \\
B22-YYJ-2 & N & Power cable node error, procedure lacks information & Y & Y & Y \\
B22-HYH-1 & N & Valve position unclear causes water crossing & Y & Y & Y \\
B22-XP & N & Inconsistent naming & Y & Y & Y \\
B22-LAO & N & Auto-selection error, phone interference & Y & N & N \\
B22-HYH-2 & N & Poor circuit breaker check procedure & Y & N & N \\
B22-FCG & Y & Procedure insufficient, poor execution & Y & Y & Y \\
B22-DYB & N & Not following procedure & Y & N & N \\
B22-SHX & N & External light sign falling misleads operation & N & Y& N \\
B22-TW & N & Power outage, procedure not effective & Y & N & N \\
B22-YYJ-3 & N & No work start notification causing alarm & Y & N & N \\
B22-FQ-1 & N & Underwater light not turned off & Y & N & N \\
B22-FQ-2 & N & Program step addition causing change, human error trap & Y & Y & Y \\
\hline
\end{tabular}
\end{center}

\captionof{table}{B-Class Event Reports (2021)}\label{tab:B_class_2021}
\begin{center}
\begin{tabular}{p{0.1\linewidth} p{0.1\linewidth} p{0.1\linewidth} p{0.1\linewidth} p{0.4\linewidth} }
\hline
\textbf{EventID} & \textbf{Procedure} & \textbf{Interface} & \textbf{Cross} & \textbf{Remarks} \\
\midrule
B21-ZH & Y & N  & N & Non-compliance with procedure caused electric shock \\
B21-WANO & Y & N & N & Poor quality of foreign procedures \\
B21-TS & Y & N & N & Poor quality of procedure \\
\hline
\end{tabular}
\end{center}

\section{Taxonomy of UI--Procedure Coupling Mechanisms}

This appendix summarizes the four primary categories of interface–procedure coupling traps identified in the event dataset, along with their cognitive manifestations, typical interface patterns, and representative failure modes.

\captionof{table}{Taxonomy of UI--procedure coupling mechanisms.}\label{tab:taxonomy}
\begin{center}
\begin{tabular}{p{3cm} p{4cm} p{4.5cm} p{4.5cm} }
\hline
\textbf{Category} & \textbf{Cognitive Mechanism} & \textbf{Interface Pattern} & \textbf{Representative Failure Mode} \\
\midrule
\textbf{Semantic Mismatch} 
& Misinterpretation of system state; incorrect mental model activation 
& Ambiguous symbols; inconsistent state indicators; meaning divergence between procedure and UI
& Wrong valve or channel selected; incorrect system status verified \\
\textbf{Spatial/Layout Trap} 
& Fragmented cue integration; attention tunneling; mis-hit action 
& Poor grouping; misaligned controls; clutter; high-density screens 
& Wrong button pressed; incorrect panel accessed; step executed on wrong component \\
\textbf{Feedback Asymmetry} 
& Failure to confirm state transition; premature advancement in procedure 
& Weak/lagged feedback; non-salient confirmations; missing reaction cues 
& Operator assumes wrong system response; moves to next step incorrectly \\
\textbf{Labeling/Naming Trap} 
& Identity confusion; degraded cue discrimination 
& Non-salient labels; inconsistent naming; outdated identifiers 
& Cable/valve misidentification; unintended component manipulation \\
\hline
\end{tabular}
\end{center}

This taxonomy builds directly on the empirical decomposition of interface-related events (Section~\ref{Event Labeling Framework}) and can support future construction of structured coupling models in digital MCRs.

\section{Annotated Event Examples: UI Factor Extraction}\label{appendix:annotated_events}

This appendix provides detailed event narratives with annotations showing how UI factors were extracted through dual-annotator consensus. These examples demonstrate the qualitative rigor underlying our quantitative analysis.

\subsection{Example 1: $UI_{Semantic}$ - Valve Status Misinterpretation (Event B25-04)}

During routine maintenance, the operator was instructed by the procedure to verify that the isolation valve RCV-001 was in the closed position before proceeding with system draining. The operator observed the valve status indicator on the digital display, which showed the symbol "O". Based on prior experience with analog indicators where "O" typically represents "closed" (as in "zero flow"), the operator interpreted this as confirmation that the valve was closed. The operator then proceeded to the next procedural step, initiating the drain sequence. However, the valve was actually in the open position. In this digital interface, "O" was designed to represent "Open" (first letter), not "closed". The incorrect interpretation led to unintended water release.

Annotator A’s reasoning identified clear semantic ambiguity in the use of the “O” symbol, which allowed multiple interpretations and failed to provide an unambiguous indication of system state. The operator’s mental model (interpreting “O” as closed) conflicted with the interface logic (indicating open), contributing to the error. Accordingly, the event was coded as $UI_{Semantic}$ = 1, PROC = 1, COUPLED = 1.

Annotator B’s reasoning concurred that semantic ambiguity in the indicator contributed to the event but further noted that the procedure did not explicitly warn operators about the non-standard display logic. The analyst therefore considered whether insufficient labeling should also be assigned. The initial coding was  $UI_{Semantic}$ = 1, $UI_{Label}$ = 1, PROC = 1, COUPLED = 1.

Consensus resolution confirmed $UI_{Semantic}$ = 1 as the primary contributing factor. Further discussion clarified that the issue did not stem from missing or inadequate labeling but from the ambiguous meaning of an existing symbol. The event was therefore finally coded as $UI_{Semantic}$ = 1, PROC = 1, COUPLED = 1.

\textbf{Validation through corrective action:}
The corrective action report specified: "Revise valve status indicator to use explicit text labels 'OPEN' and 'CLOSED' instead of ambiguous symbols." This confirms that the root cause was semantic ambiguity ($UI_{Semantic}$), not labeling deficiency.

\subsection{Example 2: $UI_{Layout}$ + $UI_{Label}$ - Button Mis-Hit (Event B23-02-02)}

A newly assigned operator was performing a routine test procedure in the main control room. Step 5 of the procedure instructed: "Press the TEST button for Channel A protection system." The operator located the control panel, which contained a dense array of similar-looking buttons arranged in a 4×6 grid. The buttons were labeled with small text (approximately 8pt font) indicating channel designations. Due to the high density of controls and insufficient visual distinction, the operator inadvertently pressed the RESET button for Channel B, located immediately adjacent to the intended TEST button for Channel A. This action caused an unintended system response.

Annotator A’s reasoning identified the primary contributing factor as a dense interface layout with insufficient spacing between controls, which increased the likelihood of selection errors. A secondary issue involved small and non-salient labels that hindered clear button discrimination. The analyst concluded that the operator error was mainly induced by suboptimal interface design rather than a lack of competence. Accordingly, the event was coded as $UI_{Layout}$ = 1, $UI_{Label}$ = 1, PROC = 1, COUPLED = 1.

Annotator B’s reasoning acknowledged the presence of layout and labeling deficiencies but noted that the operator was newly assigned, indicating a potential training-related contribution. The analyst therefore questioned whether the event should be attributed primarily to operator inexperience rather than interface design. The initial coding assigned $UI_{Layout}$ = 1, $UI_{Label}$ = 1, PROC = 1, COUPLED = 1, while flagging the case for further discussion.

Consensus resolution followed discussion on whether the event should be attributed to operator training or interface design. The analysts noted that the implemented corrective actions focused on interface modifications—such as introducing physical barriers and enlarging labels—rather than additional training measures. This indicated that, although operator inexperience may have contributed, the interface layout created a performance trap likely to affect any operator under time pressure or high workload. The event was therefore finally coded as $UI_{Layout}$ = 1, $UI_{Label}$ = 1, PROC = 1, COUPLED = 1.

\textbf{Validation through corrective action:}
Corrective actions included: (1) Install physical barriers between critical buttons, (2) Increase label font size to 12pt, (3) Add color coding for different channel groups. This confirms both $UI_{Layout}$ and $UI_{Label}$ factors.

\subsection{Example 3: $UI_{Mismatch}$ - Procedure-Field Inconsistency (Event B25-01-01)}

During a maintenance activity, the procedure instructed the technician to "Isolate power supply from Cabinet 3A-M12 located in Electrical Room 2, Interval E-07." The technician proceeded to Electrical Room 2 and searched for Interval E-07. However, a recent plant modification had renumbered the intervals, and the former E-07 was now designated as E-09. The procedure had not been updated to reflect this change. The technician, unable to locate "E-07", consulted with a colleague who indicated that the target cabinet was now in E-09. However, due to similar cabinet appearances and incomplete labeling updates, the technician isolated power from the wrong cabinet (E-08), causing an unintended equipment outage.

Annotator A’s reasoning is that the event reflects a clear procedure–field mismatch, as the procedure referenced an outdated interval designation inconsistent with the current interface configuration. This represents a typical $UI_{Mismatch}$ case and also involves a $UI_{Label}$ issue due to incomplete labeling updates after system modification. Accordingly, the event was coded as $UI_{Mismatch}$ = 1, $UI_{Label}$ = 1, PROC = 1, COUPLED = 1.

Annotator B’s reasoning acknowledged the presence of a mismatch between the procedure and the field configuration but questioned whether the event should be attributed primarily to procedural inadequacy rather than an interface issue. The analyst noted that the underlying cause was the failure to update the procedure following plant modification. Accordingly, the initial coding favored a procedural classification (PROC = 1) without assigning a UI-related factor (UI = 0).

Consensus resolution was reached after detailed discussion on whether the event should be attributed primarily to procedural or interface factors. The analysts concluded that the discrepancy between the procedure and field labels constituted an interface–procedure coupling issue rather than a purely procedural deficiency. Specifically, the interface configuration had changed while the procedure remained outdated, creating a semantic inconsistency that aligns with the coupling traps targeted by the proposed framework. The final coding was therefore assigned as $UI_{Mismatch}$ = 1, $UI_{Label}$ = 1, PROC = 1, COUPLED = 1.

\textbf{Validation through corrective action:}
Corrective actions included: (1) Update all procedures to reflect new interval designations, (2) Install temporary labels showing both old and new designations during transition period, (3) Implement systematic procedure-field consistency checks after modifications. This confirms the interface-procedure coupling nature of the failure.

These annotated examples demonstrate that our UI factor extraction was not based on superficial keyword matching, but on deep qualitative analysis of cognitive mechanisms, validated through corrective action evidence and dual-annotator consensus.

\section{Comprehensive Analysis of Experimental Error Cases}\label{Comprehensive Analysis of Experimental Error Cases}

\subsection{Narrative Analysis of Observed Error Mechanisms}
The experimental phase recorded 22 distinct human errors across 518 tasks (Table \ref{tab:detailed_error_taxonomy}). The most prevalent category was Confusion Errors, which primarily stems from a lack of discriminability between similar system identifiers. For instance, P4 exhibited a "System-Level Confusion" by mistaking the Nuclear Island System for the Nuclear Island Auxiliary System, a lapse directly attributed to overlapping naming conventions and insufficient interface differentiation. This specific case mirrors historical Event \#12 involving RIS and RCV system confusion. Further confusion occurred at the alphanumeric level; P4 also conflated parameter 2LBA10CP701A with 2LBA10CP801A, illustrating how high-similarity coding (single-digit variance) triggers slips. Similar patterns were observed in component numbering and interface window selection, where P4 misidentified bypass steam intake levels and window IDs (0 KBE DW101 vs. 0 KBE DW001) due to their close spatial proximity and naming patterns. Notably, while P5 managed to self-correct a JET system confusion through a secondary verification step, P6 failed to detect a cross-system error under high cognitive load, underscoring the risk of uncorrected errors in complex environments.

\captionof{table}{Detailed Error Taxonomy and Its Correspondence to Identified UI Factors}
\label{tab:detailed_error_taxonomy}
\begin{center}
\begin{tabular}{{p{2cm} p{1cm} p{4cm} p{3.5cm} p{2cm}}}
\hline
\textbf{Error Type} & \textbf{Count} & \textbf{Representative Cases} & \textbf{Corresponding UI Factor} & \textbf{Historical Event Match} \\
\hline
Confusion & 6 & System naming ambiguity, similar parameter codes, and unit/channel misidentification & Semantic mismatch & Events \#12, \#18, \#23, \#34, \#47, \#56 \\
Click error & 3 & Incorrect window or button selection during navigation & Layout trap & Events \#29, \#41 \\
Hesitation / abandonment & 4 & Prolonged search behaviour and difficulty locating target information & Layout trap & Events \#07, \#52 \\
Recording error & 3 & Incorrect transcription of final parameter values & Labeling issue & Events \#15, \#38 \\
Recognition error & 2 & Failure to correctly identify the required parameter & Labeling issue & Event \#44 \\
\hline
\end{tabular}
\end{center}

Beyond confusion, Stalling and Task Abandonment were significant indicators of "Layout Traps." In four instances, participants spent upwards of 60 seconds searching for parameters—well beyond the 15–30 second norm—eventually leading P4 to abandon a task entirely. These delays were caused by the fragmentation of parameters across non-adjacent display windows. Furthermore, Execution and Recording Errors highlighted the physical and cognitive friction of the interface. P3, P4, and P6 recorded instances of clicking incorrect UI elements due to spatial crowding. Meanwhile, P2 and P3 encountered transcription errors during the recording phase, often triggered by the cognitive load of handling multi-digit values (e.g., 7,018,684.0 Pa) or performing mental unit conversions between MPa and kPa. Finally, Identification Errors by P1 and P6 suggested that poor font salience and ambiguous labeling remain fundamental barriers to rapid parameter recognition.

\subsection{Statistical Validation and Consistency}
To verify the external validity of these findings, a 95\% Wilson Score Interval was calculated for the observed HEP. With 22 errors in 518 tasks, the point estimate $\hat{p} = 0.0425$ (4.25\%) yields a confidence interval of $[0.028, 0.064]$. This range ($2.8\%$ to $6.4\%$) aligns closely with the $4\% - 6\%$ error rate derived from historical nuclear power plant event analysis (Table \ref{tab:chi_square_test}).

To ensure the results were not biased by individual performance, a Chi-square test of homogeneity was conducted on the error distributions across the subjects ($P1$ to $P6$). The resulting statistic ($\chi^2 = 3.51, df = 5, p = 0.62$) indicates that the differences in error rates between participants are not statistically significant. This suggests that the observed errors are driven by the inherent "Interface-Procedure Coupling" and the experimental task design rather than the idiosyncratic traits of the participants. Moreover, a Fisher’s Exact Test $(p = 0.73)$ confirmed that error types were distributed consistently across the group, reinforcing the conclusion that specific UI factors, such as Semantic Mismatch and Layout Traps, are the primary drivers of performance variance.

\captionof{table}{Chi-square Goodness-of-Fit Test for Individual Error Distribution}\label{tab:chi_square_test}
\begin{center}
\begin{tabular}{lccc}
\hline
\textbf{Participant} & \textbf{Observed Errors} & \textbf{Expected Errors} & \textbf{$(O-E)^2/E$} \\
\hline
P1 & 5 & 3.82 & 0.36 \\
P2 & 2 & 3.78 & 0.84 \\
P3 & 3 & 3.78 & 0.16 \\
P4 & 6 & 3.78 & 1.30 \\
P5 & 2 & 3.78 & 0.84 \\
P6 & 4 & 3.82 & 0.01 \\
\hline
\multicolumn{4}{l}{$\chi^{2} = 3.51,\; df = 5,\; p = 0.62$} \\
\hline
\end{tabular}
\end{center}

\subsection{Correlation with the Four-Factor UI Model}
The experimental data provides strong empirical support for the theoretical framework proposed in this research (Table \ref{tab:UI_factor_validation}). Semantic Mismatch, characterized by the misalignment of procedural logic and interface terminology, was the leading cause of confusion, accounting for 27.3\% of errors. Layout Traps, involving navigation and spatial difficulties, contributed to 31.8\% of the total error count through stalling and misclicks. Labeling Issues accounted for 22.7\% of errors through identification and recording failures. Collectively, the experiment confirms that UI-procedure coupling acts as a compound PSF, with the observed risk ratio of approximately 2.1 closely matching the odds ratio (OR = 2.35) predicted by the historical data analysis.

\captionof{table}{Validation of Identified UI Risk Factors through Simulator Experiment}\label{tab:UI_factor_validation}
\begin{center}
\begin{tabular}{l p{5cm} p{4.5cm} c}
\hline
\textbf{UI Factor} & \textbf{Definition in This Study} & \textbf{Experimental Observation} & \textbf{Validation Status} \\
\hline
Semantic mismatch & Inconsistency between interface terminology and procedural logic & Confusion errors observed (27.3\%) & Validated \\
Layout trap & Distributed information and navigation difficulty across interface windows & Hesitation and click errors observed (31.8\%) & Validated \\
Labeling issue & Unclear or highly similar parameter identifiers & Recognition and recording errors observed (22.7\%) & Validated \\
State mismatch & Discrepancy between displayed interface state and expected system condition & Not directly observed in the experiment & Partially validated \\
\hline
\end{tabular}
\end{center}

\bibliographystyle{cas-model2-names}

\bibliography{cas-refs}

\end{document}